\documentclass[%
 reprint,
 amsmath,amssymb,
 aps,
]{revtex4-1}
\usepackage{subfig}
\usepackage{graphicx}% Include figure files
\usepackage{dcolumn}% Align table columns on decimal point
\usepackage{bm}% bold math
\begin{document}

\title{A machine learning assessment of the two states model\\for lipid bilayer phase transitions}% 
% appraisal - assessment

\author{Vivien WALTER}
\affiliation{%
 Department of Chemistry\\
 King's College London\\
 Britannia House, 7 Trinity Street, SE1 1DB, London, United Kingdom 
}%

%\collaboration{MUSO Collaboration}%\noaffiliation

\author{C\'eline RUSCHER}
% \homepage{http://www.Second.institution.edu/~Charlie.Author}
\affiliation{
Stewart Blusson Quantum Matter Institute\\
University of British Columbia\\
Vancouver BC V6T 1Z1, Canada
}%

\author{Olivier BENZERARA}
\author{Carlos M MARQUES}
\author{Fabrice THALMANN}
\email{fabrice.thalmann@ics-cnrs.unistra.fr}
\affiliation{%
 Institut Charles Sadron\\
 CNRS and University of Strasbourg\\
 23 rue du Loess, F-67034 Strasbourg, France
}%

%\collaboration{CLEO Collaboration}%\noaffiliation

\date{\today}% It is always \today, today,
             %  but any date may be explicitly specified

\begin{abstract}
We have adapted a set of classification algorithms, also known as Machine Learning, to the identification of fluid and gel domains close to the main transition of dipalmitoyl-phosphatidylcholine (DPPC) bilayers. Using atomistic molecular dynamics conformations in the low and high temperature phases as learning sets, the algorithm was trained to categorize individual lipid configurations as fluid or gel, in relation with the usual two-states phenomenological description of the lipid melting transition. We demonstrate that our machine can learn and sort lipids according to their most likely state without prior assumption regarding the nature of the order parameter of the transition. Results from our machine learning approach provides strong support in favor of a two-states model approach of membrane fluidity.
\end{abstract}

\maketitle

%\section{Introduction}

Phospholipid molecules play a major structural role in biological membranes~\cite{Mouritsen_MatterOfFat, Cevc_Marsh_PhospholipidBilayers} where a deep understanding of the physical properties can only be acquired from a detailed knowledge of the lipid assemblies. Thanks to their amphiphilic nature and geometrical characteristics, most phospholipid molecules spontaneously self-assemble in water as bilayers. Supported or free-standing lipid bilayers and vesicles can easily be made, controlled and studied, and have now become standard tools in membrane biophysics studies, referred as  model lipid bilayer systems~\cite{Dimova_Marques_TheGiantVesicleBook}. Early studies on pure phospholipid bilayers indicated that lipids were subject to thermodynamic transitions~\cite{1976_Mabrey_Sturtevant,Cevc_Marsh_PhospholipidBilayers,Heimburg_BiophysicsMembrane,Marsh_HandbookLipidBilayers2}, with in particular a sharp transition associated to a significant change in enthalpy called \textit{main}, or \textit{melting} transition. This transition separates a low temperature well-packed assembly from a high temperature expanded, disordered lipid tail organization. This transition is considered to be of first order, although phase coexistence is difficult to establish experimentally \cite{2012_Armstrong_Rheinstadter}. The low temperature phase is commonly referred as the \textit{gel} state, while the high temperature state is the \textit{fluid} state. Lipid mixtures also display melting transitions spreading along a finite temperature range, usually accompanied by gel-fluid domain coexistence. It is usually assumed that most biological lipid membranes are found in a fluid state, and many scenarii aiming at explaining the lateral lipid and protein segregation observed in biological membranes involve ordering of the lipid tails.  

The consensual description of the single component lipid melting transition assumes that dominant molecular conformations evolve from  all-\textit{trans} extended, well oriented hydrocarbon chain conformations in the low temperature phase, to disordered chains melted by rotation isomerism, as proposed in the earliest theoretical proposals~\cite{1973_Nagle, 1974_Marsh, 1974_Marcelja_2, 1979_Pink_Chapman}. Lipids in the fluid phase have more configuration entropy and more enthalpy than those in the gel phase, due to lower density and cohesive energy, and higher chain torsion energy. Balance between entropy and enthalpy holds at the melting temperature $T_m$. Despite some asymmetry between the two phases, a large number of experimental facts related to melting transition of pure and mixed lipid compositions have been successfully interpreted by means of a phenomenological two-states model, originally proposed by Doniach~\cite{1978_Doniach, 1983_Sugar_Monticelli,1996_Heimburg_Biltonen,1996_Jerala_Biltonen, 1999_Sugar_Biltonen, 2001_Ivanova_Heimburg, Heimburg_BiophysicsMembrane, 2011_Wolff_Thalmann, 2018_Morandi_Marques}. This model can be expressed as an Ising model, each lipid taking binary discrete values (say $s=0$ for \textit{gel} and $s=1$ for \textit{fluid}) with neighboring lipids being coupled~\cite{Heimburg_BiophysicsMembrane}. In the framework of the two-states model, the Ising variables stand for a coarse-grained description of the lipid tail conformations, assuming that lipids can be classified into two classes, according to their molecular conformations. Within this description, an effective temperature dependent "magnetic field" $h(T)$ biases the odds in favor of one or the other state, while cooperativity results from nearest neighbor state coupling. 

Usually, the determination of the lipid bilayer state relies on a structural scalar order parameter, such as the membrane thickness, given for instance by the head to tail lipid extension or the tail molecular order parameter. We address in this work the validity of the two-states description for the transition of pure DPPC  (1,2-dipalmitoyl-\textit{sn}-glycero-3-phosphatidylcholine) bilayers, using atomistic molecular dynamics (MD) simulations and supervised Machine Learning (ML) classification algorithms. Assessing the two states model requires ones to analyze single lipids and sort them into their respective states. Our ML classification works without reference to an existing or newly defined scalar order parameter. It only relies on a procedure for lifting the orientation degeneracy of each lipid configuration. In addition, the ML approach may serve as testing the relevance of a given scalar order parameter \textit{a posteriori}. 

Machine Learning has already been applied successfully to a number of situations in statistical thermodynamics and phase transitions. For instance, Cubuk \textit{et al.} used support vector machines to localize plastic flow regions in amorphous structures~\cite{2015_Cubuk_Liu}, Carrasquilla and Melko revealed the strong aptitude of neural networks models to recognize various spin ordering regimes in condensed matter systems~\cite{2017_Carrasquilla_Melko}, Le and Tran succeeded in predicting the polymorphism of complex lipid mixtures given a set of structural, chemical and composition parameters, by means of an artificial neural network approach~\cite{2019_Le_Tran}. We show in the present work that Machine Learning classification tools are also extremely powerful for investigating thermal lipid phase transitions.

%investigation of thermodynamics and phase transitions is still an active field of research in condensed matter where ML has been shown to play a significant role.

% and is a very active field of research
%********************************************************************************  %
\begin{figure}[ht!]
\centering{
\includegraphics[scale=.37]{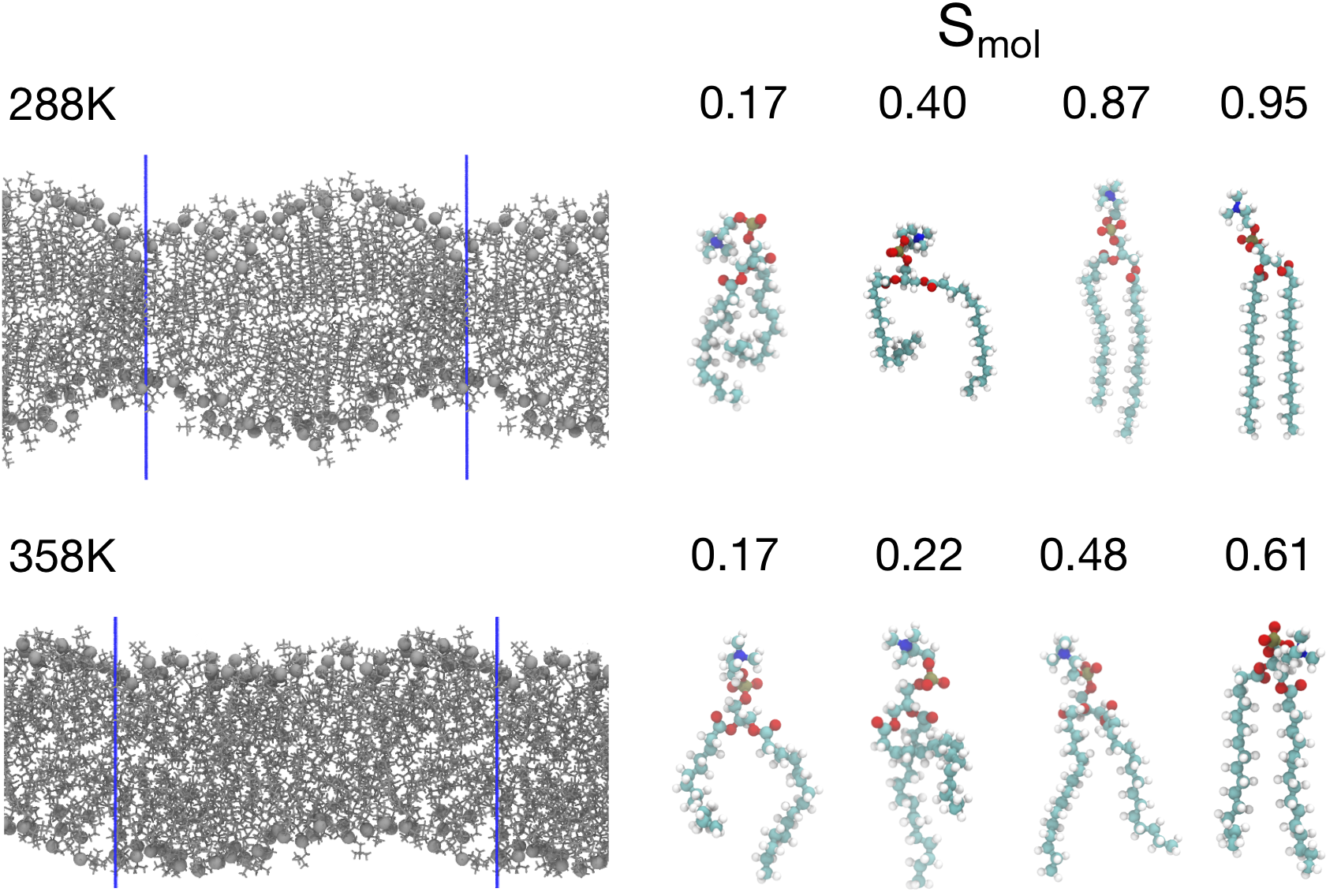}
\caption{Snapshots of a DPPC bilayer simulated at (top-left) 288~K and at (bottom-left) 358~K, respectively below and above the experimental $T_m$ of the lipid. Lipid molecules are shown in gray, with their phosphorus atom displayed as a plain big sphere to distinguish the lipid orientation in the bilayer. Blue lines delimit the simulation box beyond which periodic boundary conditions are applied to the system. Right: lipids extracted from the bilayers displaying an important diversity in their conformations, with their associated molecular order parameter $S_{\mathrm{mol}}$. Top-right: conformations at 288K, bottom right: conformations at 358K.}
\label{fig_system}
}
\end{figure}
% ******************************************************************************** %

DPPC molecules are among the best known phospholipids~\cite{Marsh_HandbookLipidBilayers2}. They display experimentally  a melting transition at 314~K, which is well reproduced by the CHARMM-36/TIP3P force-field for atomistic simulations of lipid bilayers in aqueous solutions~\cite{2010_Klauda_Pastor,best_mackerell_2012}.  In our simulations, in agreement with experiments, the bilayer was found in a $L_{\alpha}$ fluid disordered state above the main transition, and in a $P_{\beta'}$ rippled phase (Figure~\ref{fig_system}), \textit{i.e} a spatially modulated gel phase with peristaltic variations in the bilayer thickness and normal direction~\cite{2005_deVries_Marrink,2018_Khakbaz_Klauda} at lower temperatures. 
Unlike Khabkaz and Klauda, our simulations did not show evidence of a $L_{\beta'}$ tilted gel phase at temperatures below the pretransition, which is expected to take place at $T= 307$~K on experimental grounds. This is likely to be due to a difference in size in our system, comprising nearly 3 times as many lipids as the one used in ref~\cite{2018_Khakbaz_Klauda}. The presence of a ripple phase, whether stable or metastable, renders approaches based on scalar order parameters impractical, while our ML approach was found to work efficiently also in this situation. In what follows, we still name our low temperature state a gel phase, in spite of its ripple character.   

The principle of the analysis is as follows. A 212 lipid molecules system was thermalized at low (288~K) and high temperature (358~K) and pressurized with a semi-isotropic barostat. At such temperatures, we assume that lipid conformations are predominantly  gel and fluid respectively. Our training set was therefore composed of an equal number of conformations coming from the 288~K MD trajectory (all labelled as gel) and the 358~K trajectory (labelled as fluid). Details regarding the numerical simulation procedure and machine learning implementation details are given as Supplemental Material (SM).

Raw molecular conformations (spatial coordinates) were processed to remove all translation and rotation degeneracy, and the dimension of the initial conformation space was slightly reduced, resulting respectively in a 100-dimensional reduced coordinate space $\mathcal{X}$, and a 61-dimensional mutual distances space $\mathcal{D}$. The coordinate space $\mathcal{X}$ and the distance space $\mathcal{D}$ provide a very detailed description of individual lipid conformations, and constitute the starting point of the ML approach. They are referred as "feature spaces" in the context of automated classification (see SM). The processed conformations were then fed to the ML algorithms. Three different algorithms were selected for the purpose of classifying the  molecular conformations: Naive Bayes (NB), K-Nearest Neighbors (KNN) and Support Vector Machines (SVM). They were all used as implemented in the \textit{Python/Scikit-learn} package~\cite{pedregosa_duchesnay_2011}. The different ML algorithms were tested and found to perform moderately well when used separately, some methods performing better for gel lipids, other methods for fluid lipids. Combining the approaches together, we managed to get a success rate of 88\% upon validation, \textit{i.e.} using 80\% of the training set configurations for learning, our best ML algorithm was able to assign 86\% of the low temperature configurations to a gel state, and 91\% of the high temperature configurations to a fluid state.  After training, the ML model was used to analyse simulations at arbitrary temperatures. 

One important prediction of the two states model is the presence of minor components in the majority phase, under the form of "thermal excitations". A finite fractions of molecules tend to adopt a conformation different from their immediate environment, in spite of the presence of a "local field" biasing the statistics in favor of the dominant state. In that respect, the training set cannot be considered as containing only pure gel and fluid conformations. The presence of minor components is inherent to the presence of thermal excitations, and it does not seem possible to curate the training set without introducing further unwanted biases into the analysis. Our results show, however, that the training algorithm is not sensitive to the presence of a small fraction of non representative lipid conformations. In other words, the learning procedure was found to be robust so long as the training sets temperatures were chosen far apart from the melting transition. 

\begin{figure}[ht!]
\centering{
\includegraphics[scale=.35]{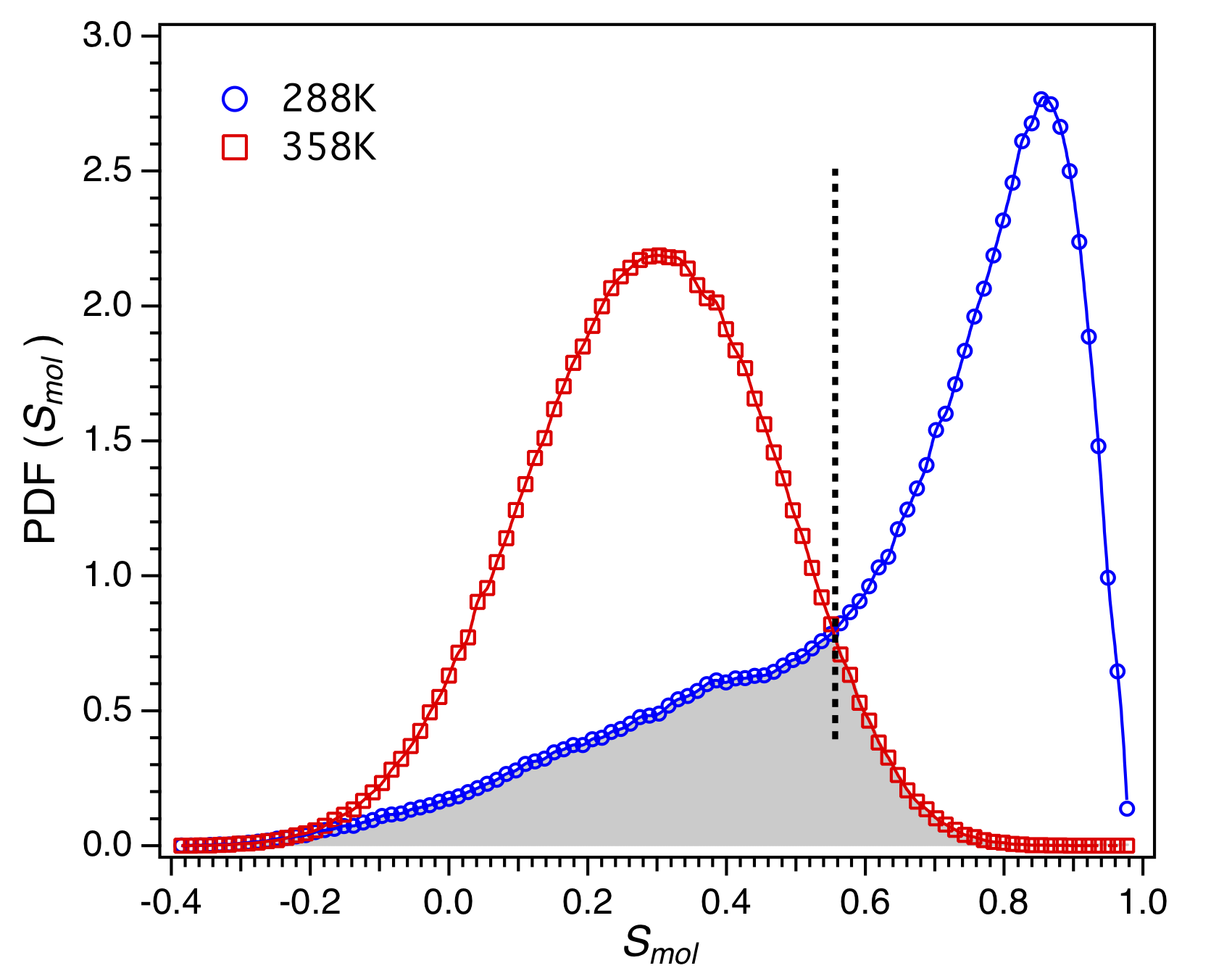}
\caption{Distribution of the molecular order parameters $S_{\mathrm{mol}}$ obtained from MD trajectories at 288 and 358~K respectively (using $10^6$ lipid conformations). The dashed black line shows the threshold value that should be set to best determine the internal lipid states. The fraction of the population which would be incorrectly classified by using this threshold is represented by the shaded area in gray.}
\label{fig_order_parameter}
}
\end{figure}

Let us first analyze how ML predictions differ from those based on standard scalar structural order parameters. Figure~(\ref{fig_order_parameter}) shows the distribution of the molecular segmental order parameter $S_{\mathrm{mol}}$, below and above the transition. This scalar observable was defined, for a given lipid and an instantaneous configuration (coming from a MD trajectory frame), by averaging over all the CC bonds in the two aliphatic tails a nematic parameter $(3\cos(\theta)^2-1)/2$, $\theta$ being the bond orientation with respect to the bilayer normal direction $z$. Both histograms overlap significantly. Using for instance a threshold value $S_{\mathrm{th}} =0.55$, where the estimated probability density functions (pdf) coincide, it was found that altogether 17\% of lipids (1 in 6) were set to be assigned to the opposite state, either gel at 358~K, or fluid at 288~K. The gel $S_{\mathrm{mol}}$ distribution appears to be strongly skewed, as 28\% (1 in 3) of the lipid molecules end up in the fluid state at 288~K. The relatively large fraction of lipids with $S_{\mathrm{mol}}\leq 0.55$ found at 288~K is in part due to the ripple structure of the bilayer. This clearly show that using $S_{\mathrm{mol}}$ for categorizing lipid conformations gives, at best, unreliable results. A similar analysis was conducted for two other scalar determinants: the lipid head-to-tail extension $L$, and the area per lipid $A_l$ computed from a Voronoi tessellation of the 2d $xy$  projection of lipid center of mass positions (assigning to each lipid a polygonal cell of given area). Both were found to perform worse than the segment orientation $S_{\mathrm{mol}}$ approach (see SM). 

% *************************************************** %
% *************************************************** %

\begin{figure}[!ht]
\centering{
\includegraphics[scale=.35]{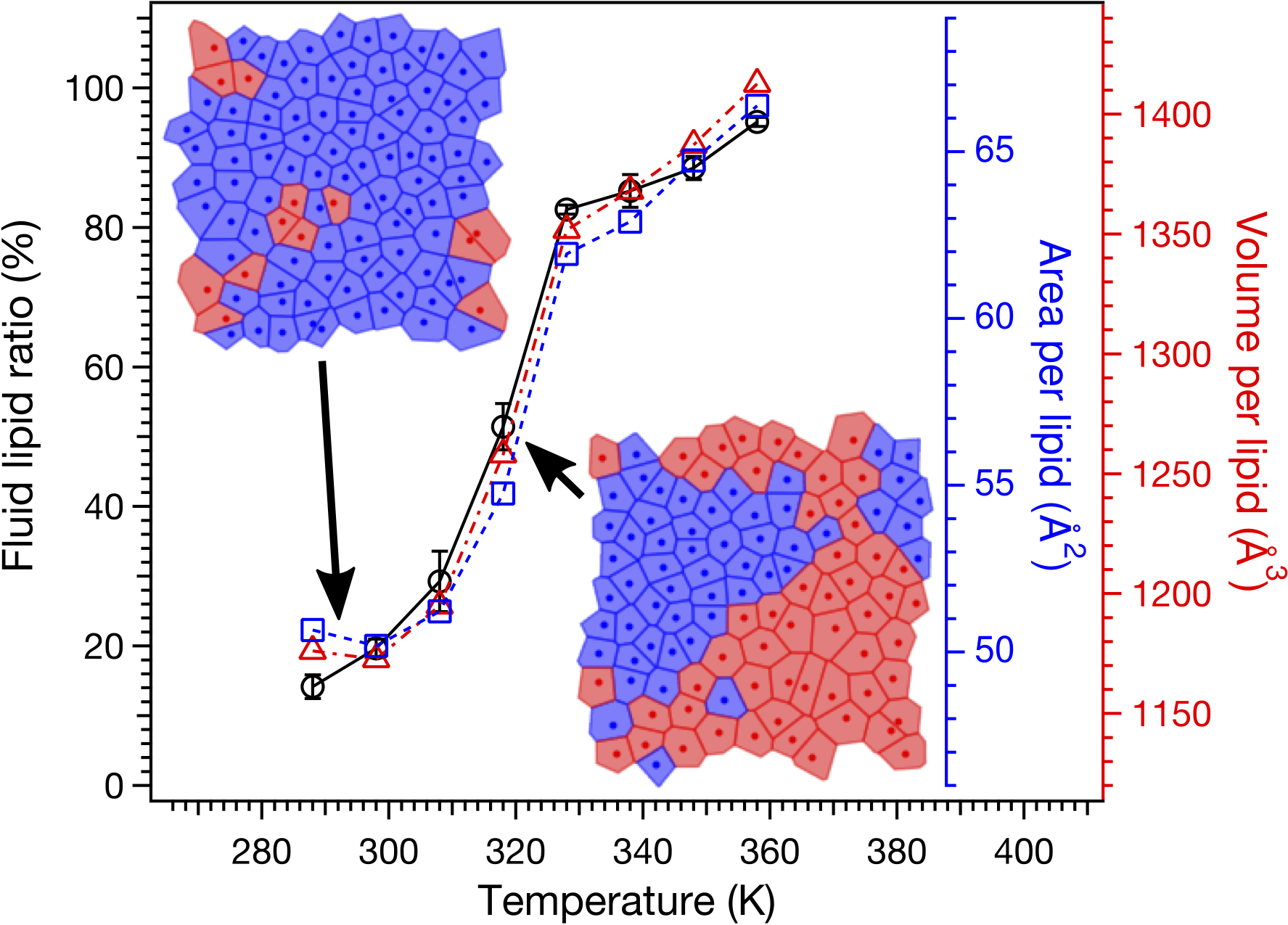}
\caption{Ratio of lipids in fluid state (black curve, vertical axis on the left), area per lipid (blue curve, vertical axis on the right) and volume per lipid (red curve, vertical axis on the right) as function of temperature. Two snapshots of the upper membrane leaflet, respectively taken at 288 and 318~K, are shown in the inset. The dark spots correspond to the 2d projection of the lipid center of masses. Boundaries between lipids result from the associated 2d Voronoi tessellation. Each cell color corresponds to the assignment of the ML, blue for the gel state and red for the fluid state.}
\label{fig_phase_transition}
}
\end{figure}

To provide further evidence in favor of our approach, we compared the ML predictions to the  membrane structure as a function of temperature. Figure~(\ref{fig_phase_transition}) shows on the same graph the average area per lipid $A_l$, the average volume per lipid $V_l$ and the ML prediction for the fraction of lipid assigned to the fluid state. The area per lipid is defined as the projected membrane area divided by the number of lipids per leaflet, neglecting out-of-plane membrane undulations. The volume per lipid results from a Voronoi analysis of the lipid center of masses. The structural values evolve monotonically and reversibly with temperature, from 288 to 358~K. The three curves superimposes very well, which shows that structural data supports the finding of our classification tool. Two snapshots of the membrane upper leaflet are also provided, at low and intermediate temperature. The low temperature membrane shows a small number of isolated small clusters of fluid lipids. A one to one fluid/gel ratio is observed at 318~K (51 $\pm$ 3\% of lipids in the fluid phase), a temperature close to the experimental melting temperature (314~K).
%314~K than the value of 321~K found by fitting the evolution of the area per lipid 
A snapshot of the upper leaflet at 318~K shows two large distinct domains, separated by a smooth boundary.  The fluid state ratio evolves smoothly and reversibly from an estimated value of 14 $\pm$ 2\% at 288~K to  95 $\pm$ 1\% at 358~K. The very strong correlation between fluid ratio $x_f$ and area per lipid $A_l$ can be naturally interpreted in the framework of the two-states model by assigning constant values $A_f$ and $A_g$ to the fluid and gel states, resulting in $A_l=x_f A_f+(1-x_f)A_g$. A similar conclusion could be reached when considering the volume per lipid $V_l$.

% ************************************************************************ %
% ************************************************************************* %

The ML predictions agree well with the global structural properties of the membrane and we now consider the local correlation properties. The two-states model combines the internal spontaneous dynamics of each lipid with local interactions promoting cooperativity between neighbors. The interactions, known as $J$ coupling in the Ising model context, are responsible for the sharp structural and thermodynamic changes with temperature, the emergence of local correlations, clustering and domain formation, such as seen in Fig~\ref{fig_phase_transition}. Such couplings create a local field whose effect is to bias the lipid state in favor of the dominant local phase, according to the majority rule. Internal reversible gel $\leftrightarrow$ fluid state transitions occur spontaneously, according to a non conserved parameter, or Glauber dynamics~\cite{1963_Glauber}. In order to get insight into the local correlations and flip rates, we performed a systematic statistical neighbor analysis, and estimated the conditional conversion states.  

\begin{figure}[!ht]
\centering{
\includegraphics[scale=.21]{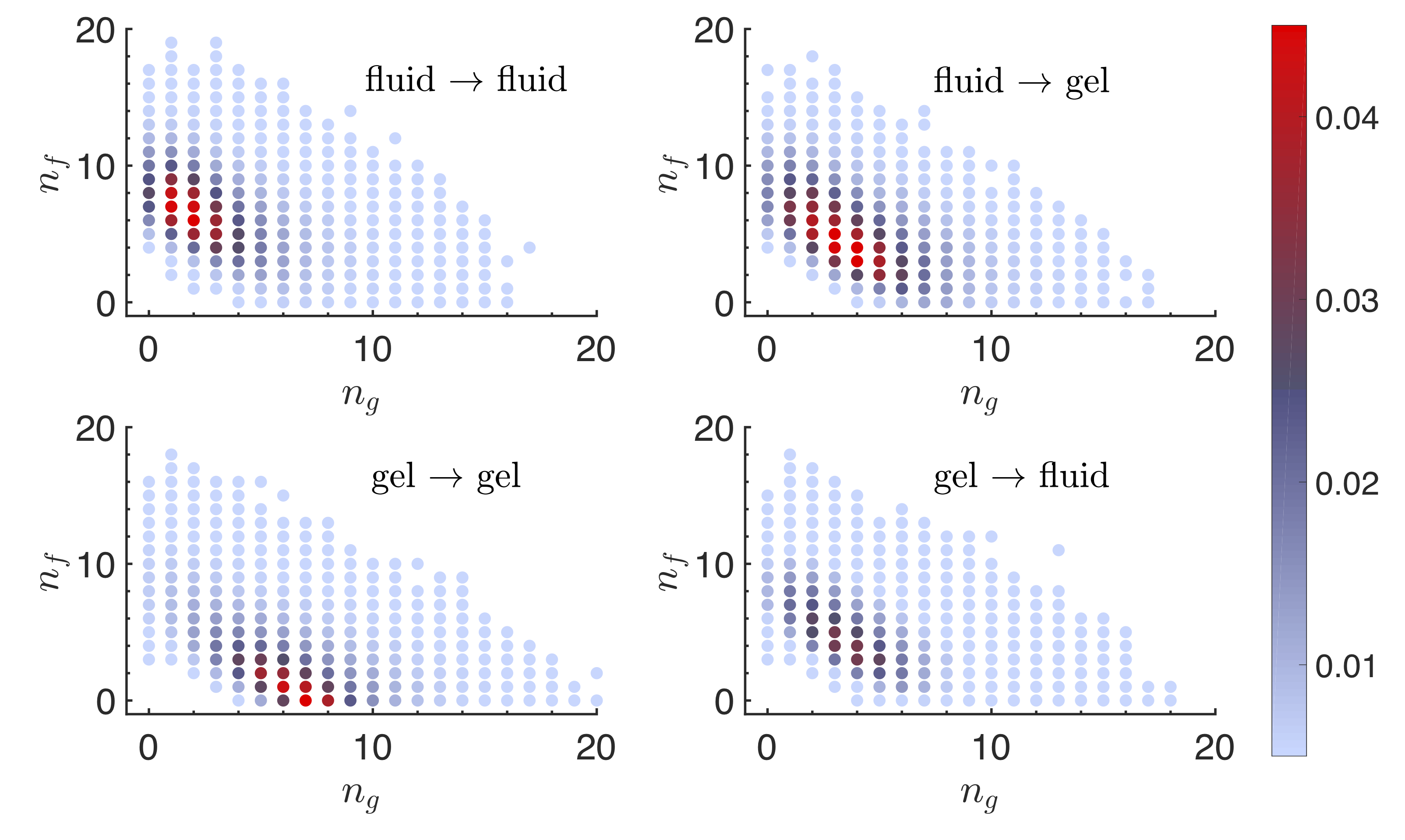}
\caption{Color coded histograms of the $(n_l,n_g)$ distribution for the 4 subsets described in the text, at $T=318~K$. Probability at $T=318K$ for a lipid to have its local environment composed of $n_g$ neighbors in gel state and $n_f$ neighbors in fluid state. Left column informs about the local environment probability of lipids in the fluid state (Top) and gel state (Bottom) when no change in the state is observed.   Right column shows the probability of the local environment just before lipids experience a transition to the other state.}
\label{fig_neighbours}
}
\end{figure}

Voronoi tessellations provide an operational method for deciding unambiguously which lipid pairs are nearest neighbors. Following a protocol described in the SM, we consider two consecutive MD frames and divide the lipid population into four subsets: (a) lipids categorized as fluid, which stay in the fluid state, (b)  lipids categorized as fluid, switching to the gel state, (c) lipids in the gel state, remaining in the gel state and (d) lipids in the gel state switching to a fluid state. At $T=318~\mathrm{K}$ the fluid and gel phases compete evenly and the fluid and gel populations are roughly equal. We count for each molecule, the number $n_g$ of gel neighbors and $n_l$ of fluid neighbors. We find in these conditions that the most typical environment of a lipid in fluid state (a and b) is $n_g = 2 \pm 1$ and $n_f = 7 \pm 1$, while for a lipid in gel state (c and d) $n_g = 7 \pm 1$ and $n_f = 1 \pm 1$. 
However, the internal lipid state fluctuate spontaneously, and clearly the dynamics is strongly influenced by the local environment, as demonstrated in Figure~(\ref{fig_neighbours}). Indeed,  for gel to fluid (case d) and fluid to gel (case b) transitions, we notice that all lipids that are just about to switch are more likely to have an equal number of gel and fluid neighbors (typically $n_g = 3 \pm 1$ and $n_f = 5 \pm 1$). We conclude that lipids subject to internal state transitions are mostly located at the border between domains. The results of our neighbor analysis clearly support the idea that internal state dynamics is under the control of some local field.

% *********************************************************************** %
% ********************************************************************** %

As a summary, we trained a Machine Learning algorithm to classify phospholipid molecular conformations obtained by atomistic molecular dynamics simulations. Lipids were sorted into two classes, gel and fluid, according to their similarity with a reference (training) set of conformations originating from low and high temperature trajectories respectively. The efficiency of the ML approach is superior to simple schemes based on scalar order parameters, and deals successfully with the ripple structure at low temperature. The measured fraction of fluid/gel conformations correlates very well with the observed  structural changes as temperature evolves.  A finite fraction of the minor phase is always present which can be associated to thermal excitations in the framework of a two-states Ising model interpretation. Lipids with similar state tend to cluster into large domains, while spontaneous internal state conversions are more likely to occur at the boundary between domains. The local distribution of neighbors supports the concept of local field and nearest neighbor coupling. We overall conclude that our ML approach provides convincing evidence in favor of the two-states phenomenological model. 

We foresee numerous applications of the present approach. A first straightforward extension concerns lipid binary mixtures~\cite{1988_Ipsen_Mouritsen} and liquid ordered phases caused by the presence of cholesterol~\cite{1987_Ipsen_Zuckermann}, as well as membranes of more complex composition~\cite{2014_Ingolfsson_Marrink}. We also anticipate that our ML approach will be useful to study the influence of membrane solutes that are known to influence the thermodynamics of melting in model membranes. This includes hydrophobic pollutants, \textit{e.g.} pyrene~\cite{2014_Franova_Ollila}, carbohydrates~\cite{1991_Crowe_Crowe,2018_Morandi_Marques} and synthetic oligomers~\cite{2014_Rossi_Monticelli,2014_Rossi_Monticelli_2}. A ML approach could then quantify the lipid state alteration induced by these compounds. Importantly, this analysis is also well-suited to study the local lipid environment of membrane proteins, for which the existence of lipid mediated interactions and minor phase nucleation is speculated~\cite{1997_Gil_Mouritsen,1998_Gil_Zuckermann}. For all these cases, significant improvements over  approaches relying on scalar order parameters can be anticipated.

V.W. warmly thanks Tiago Espinosa de Oliveira for helping with simulations set-up. The authors gratefully acknowledge support from the high performance cluster (HPC) Equip@Meso from the University of Strasbourg, through grant n$^{\circ}$ G2018A53.  

%\bibliographystyle{unsrt}
%\bibliography{references}

\widetext
\begin{center}
\textbf{\large Supplementary Information}
\end{center}

\setcounter{equation}{0}
\setcounter{figure}{0}
\setcounter{table}{0}
\setcounter{page}{1}
\makeatletter

\renewcommand\thefigure{S\arabic{figure}}

\newcommand{\beq}{\begin{equation}}
\newcommand{\eeq}{\end{equation}}

\newcommand{\NX}{N_{\mathcal{X}}}
\newcommand{\ND}{N_{\mathcal{D}}}

\section{Simulations}

All simulations were performed using GROMACS 2016.4~\cite{berendsen_vandrunen_1995,abraham_lindahl_2015} along with the CHARMM-36 all-atom force-field~\cite{best_mackerell_2012} (June 2015 version). A lipid bilayer made of 106 lipid molecules per leaflet, each containing 130 explicit atoms, was created using CHARMM-GUI~\cite{jo_im_2008, wu_im_2014, jo_im_2009, jo_im_2007}. It was hydrated with two 8~nm thick water layers on each side (connected through periodic boundary conditions), using the TIP3P water model, for a total of 29826 solvent molecules. The force field parameters for DPPC molecules were provided directly by CHARMM-GUI~\cite{brooks_karplus_2009, lee_im_2016}.

The above system was first subject to energy relaxation using steepest descent energy minimization, followed by a 10~ps NVT thermalization stage at 288~K. Then, the bilayer was subject to a 1~ns NPT equilibration run coupled to a semi-isotropic barostat (1~bar in all directions). The system was then further  equilibrated at the desired temperature with the same semi-isotropic barostat during a second NPT equilibration step of 10~ns. Molecular dynamics production runs of 50~ns were finally generated at the same temperature and with the same semi-isotropic barostat. The analysis were performed on the last 25~ns of simulations. All time steps were set to 2~fs.

All the molecular dynamics simulations used the leap-frog integration algorithm~\cite{hockney_eastwood_1974}. Temperature and pressure were kept constant using respectively a Nos\'e-Hoover thermostat~\cite{nose_1984, hoover_1985} (correlation time $\tau_T = 0.4$~ps) and a Parrinello-Rahman semi-isotropic barostat~\cite{nose_klein_1983, parrinello_rahman_1998} (correlation time  $\tau_P = 2.0$~ps, compressibility $4.5 \times 10^{-5}$~bar$^{-1}$). 

Lipid and water molecules were separately coupled to the thermostat. Following GROMACS recommendations for the CHARMM-36 all-atom force field, a Verlet cut-off scheme on grid cells was used with a distance of 1.2~nm, and non-bonded interactions cut-offs (Van der Waals and Coulombic) were also set to 1.2~nm. Fast smooth Particle-Mesh Ewald electrostatics was selected for handling the Coulombic interactions, with a grid spacing of 4~nm. A standard cut-off scheme with a force-switch smooth modifier at 1.0~nm was applied to the Van der Waals interactions. We did not account for long range energy and pressure corrections, and constrained all the hydrogen bonds of the system using the LINCS algorithm.

%%%%%%%%%%%%%%%%%%%%%%%%%%%%%%
%%% System Analysis
%%%%%%%%%%%%%%%%%%%%%%%%%%%%%%
\section{System Analysis}

\subsection{Determination of structural parameters}

Values of the average area per lipid $A_l$ and order parameter $S_{\mathrm{mol}}$ of the bilayer were respectively obtained using the GROMACS built-in commands \texttt{gmx energy} and \texttt{gmx order}.

\begin{figure}[h!]
\centering{
\includegraphics[scale=.2]{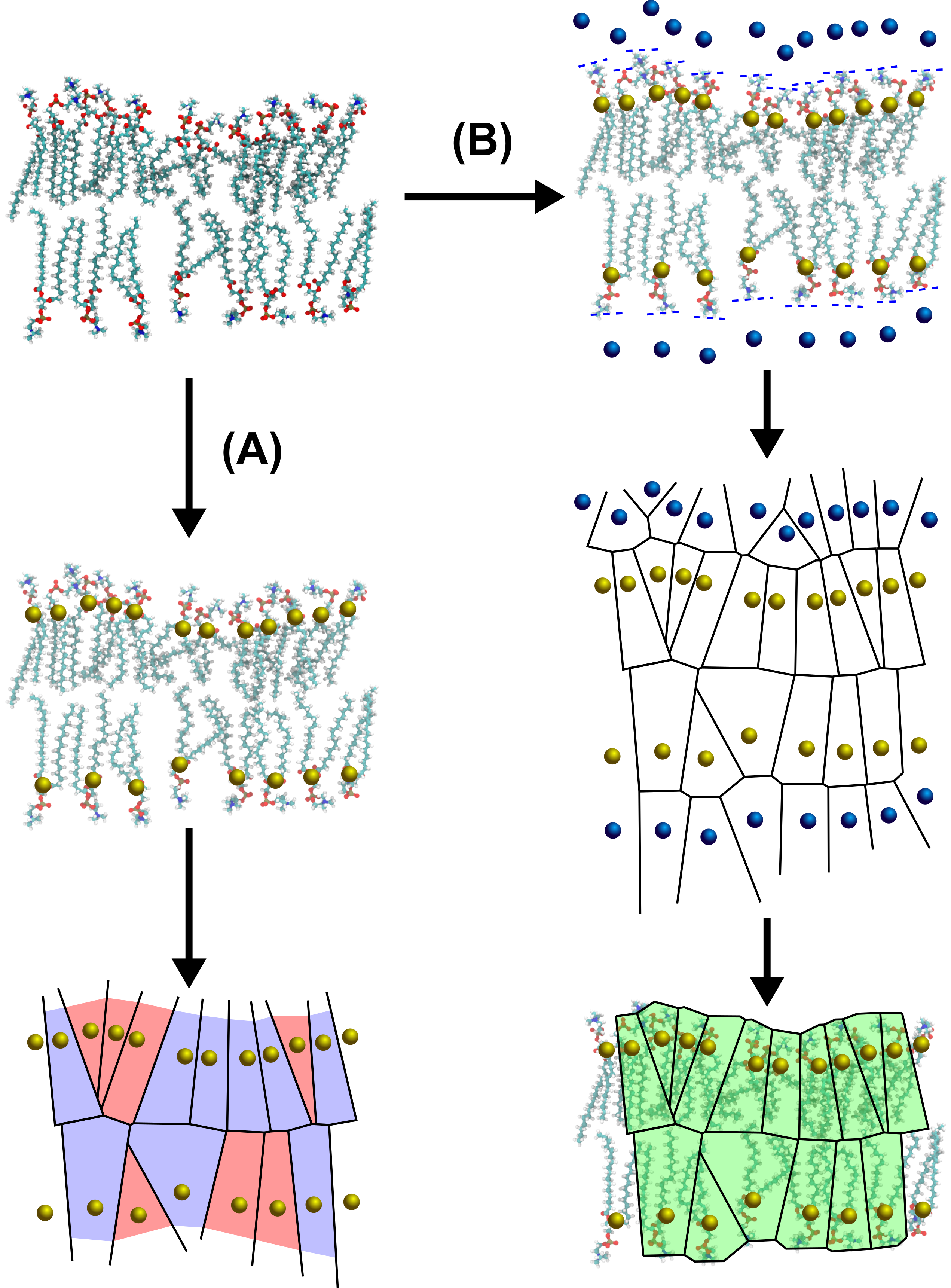}
\caption{Comparison of  3-dimensional Voronoi tessellations of a lipid bilayer configuration (A)~without and (B)~with ghost lipids. Without ghost lipids, most cells are unbounded, with infinite volume, due to the absence of particle on the opposite side of the water-membrane interface. As a practical solution of this problem, ghost lipids are added to the data set, mirroring the lipid center of mass positions. The resulting cells for lipids inside the bilayer display a realistic volume and shape, accounting for the water interface in a natural way.}
\label{sfig_ghost_lipid}
}
\end{figure}

For measurements of individual lipid properties (area, order parameter, elongation), atom positions were collected from trajectories using the Python~3 \texttt{MDAnalysis} module~\cite{gowers_beckstein_2016, michaud_beckstein_2011}. The individual areas per lipid were obtained from Voronoi tessellations of the two-dimensional projections of the lipid center of masses, computed using the \texttt{Voro++} library~\cite{rycroft_2009}. The individual volumes per lipid were derived from three-dimensional tessellations of the lipid centers of mass, again using the \texttt{Voro++} library. Note that the bilayer geometry requires a specific tessellation procedure: this was done by introducing \textit{ghosts} lipids in the water regions. Without these ghost lipids, the tessellation cells cannot be correctly defined and are unbounded across the membrane-water interface, thus overestimating significantly the individual volume per lipid. Ghosts lipids are mirror images of bilayer lipids across the local lipid-water interface (\textit{cf.} Fig.~\ref{sfig_ghost_lipid}). After the tessellation was made, ghost lipids described in the previous section and their corresponding cells were removed the lists, and only the volumes of physical lipids were collected and analyzed.

The molecular order parameter $S_{\mathrm{mol}}$ of individual lipids was calculated by measuring, for every $N_C-2=14$ non-terminal carbon atoms $k=2\ldots 15$ of the 2 tails of the lipids, the angle formed between the $z$-axis of the system directed along $\vec{u}_z$ and the vector $\overrightarrow{C_{(j,k-1)}C_{(j,k+1)}}$ defined by the carbon atoms surrounding atom $k$ within the same tail $j=1,2$. The order parameter $S_{\mathrm{mol},(j,k)}$ of the atom $k$ is obtained from the 2$^{nd}$ Legendre polynomial $P_2$ using $\cos(\theta_{(j,k)})= \vec{u\vphantom{c}}_z\cdot \overrightarrow{C_{(j,k-1)}C_{(j,k+1)}}$, and averaging over $j$ and $k$:
\begin{equation}
    \label{eq_avg_smol}
    S_{\mathrm{mol}} = \frac{1}{2(N_C-2)} \sum_{j=1}^{2} \sum_{k=2}^{N_C-1} \frac{1}{2}\left(3\cos(\theta_{(j,k)})^2-1\right)
\end{equation}

%%%%%%%%%%%%%%%%%%%%%%%%%%%%%%%%%%%%%%%%%%%%%%%%%%%%%%%%%%%%%%%

%%%%%%%%%%%%%%%%%%%%%%%%%%%%%%%%%%%%%%%%%%%%%%
\subsection{Next-nearest neighbors statistics}

After completion of the 3d Voronoi tessellation using \texttt{Voro++}, a list of next-nearest neighbors was established for each lipid center of mass. We also collected the areas of the polygonal surfaces separating each pair of neighboring Voronoi cells. The ghost lipids and their corresponding faces were removed from the lists. The neighbor lists were further curated by removing all the neighbor pairs for which the corresponding face area accounted for less than 1\% of the total surface area of each Voronoi cell in contact. The number of next-nearest neighbors were finally counted to build the coordination statistics $(n_g,n_f)$, where each lipid molecule has $n_g$ gel and $n_f$ fluid neighbors.

%%%%%%%%%%%%%%%%%%%%%%%%%%%%%%%%%%%%%%%%%%%%%%%%%%%%%%%%%%%
%%%% Machine Learning discrimination of lipid states
%%%%%%%%%%%%%%%%%%%%%%%%%%%%%%%%%%%%%%%%%%%%%%%%%%%%%%%%%%%%
\section{Machine Learning discrimination of lipid states}

%%%%%%%%%%%%%%%%%%%%%%%%%%%%%%%%%%%%%%%%%%%
\subsection{Lipid classification procedure}

The lipid classification process involves three steps:

\begin{itemize}
\item The molecular conformation of a lipid is recorded as a list of atom positions. Lipids are then shifted and rotated in order to remove all rotation and translation degeneracy.

\item The lipid molecular conformations are further simplified, reducing each single lipid conformation to a set of 100 $(r,z)$ coordinates (configuration space $\mathcal{X}$), or 61 mutual distances (configuration space $\mathcal{D}$, see below for details).

\item 424 conformations were selected with the purpose of training the algorithms. A number of \textit{Machine Learning} procedures were tested and compared. A combination of 4 algorithms was found to maximize the training success rate. We therefore combine the 4 algorithms in our final classification procedure (see below for details). 

\end{itemize}

%%%%%%%%%%%%%%%%%%%%%%%%%%%%%%%%%%%%%%%%%%%%%%%%%%%%%%%
\begin{figure}[h!]
\centering{
\includegraphics[scale=.6]{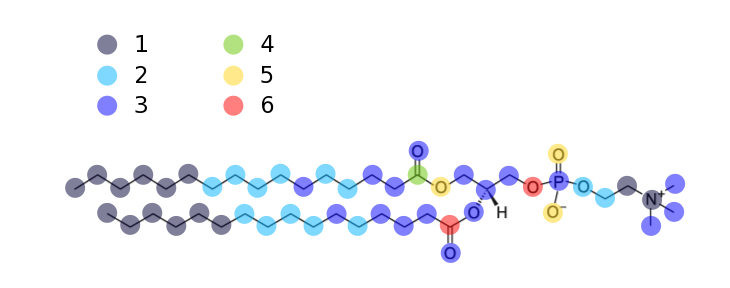}
\caption{Skeletal structure of a DPPC molecule. The atoms are colored according to the number of pairs formed between this atom and other atoms separated by exactly 6 covalent bonds along the molecule chain. The highest number of neighbors for a single atom was found to be 6. The total number of pairs is 61.}
\label{sfig_map_neighbors}
}
\end{figure}

%%%%%%%%%%%%%%%%%%%%%%%%%%%%%%%%%%%%%%%%%%%%%%%%%%%
\subsection{Definition of the configuration spaces}

Machine Learning algorithms share the ability to discriminate well multidimensional data. Our purpose is to feed the algorithms with single lipid conformations and have them sorted into two classes. We observe that the raw chemical formula of DPPC (CAS 2644-64-6) is PNO$_8$C$_{40}$H$_{80}$. Disregarding hydrogens, each lipid molecule comprises 50 "heavy atoms", and therefore the associated single lipid configurations belong to a 150 dimensions vector space. As some configurations can be mapped onto each other by means of a spatial displacement (translation and/or rotation), the set of configurations has only 144 independent degrees of freedom. 

For simplicity, we decided to work with slightly smaller configuration spaces, by projecting further the 144 dimensional molecular conformations onto two configuration spaces, $\mathcal{X}$ and $\mathcal{D}$, that we now define. Within the first approach, one determines the proper inertial frame of each lipid configuration, locating the center of mass, fitting the position of all atoms with a 3D line to find the longest axis. Lipid configurations can be recast into the inertial frame, using the center of mass as origin, and the longest axis (smaller moment of inertia) as vertical axis. This almost always results in having the lipid directed along the bilayer normal, with two possible orientations. When necessary, the lipid is flipped in order to place by convention the phosphocholine group into the positive $z$ upper half space. This combination of reorientations lifts entirely the translation and orientation degeneracy of the original coordinate space.   
New cylindrical coordinates $\{r_i,\theta_i, z_i\}$, $i=1\ldots 50$ can be associated to the resulting lipid conformation. We decided then to disregard entirely the angles $\theta_i$, keeping only the coordinate subset $\{X_i = r_i,z_i\}$, $i=1\ldots 50$. This defines a $\NX=100$ dimensional space $\mathcal{X}$ that will subsequently be referred as "Coordinate space". 

For the second approach, starting from the original 150 lipid coordinates, we calculate a set of mutual euclidean distances between pairs of atoms. Two atoms participate in a pair when they are separated by 6 positions along the chemical graph defining the molecule (see Fig~\ref{sfig_map_neighbors}). Enumerating the possibilities of the tree-like graph, one finds 61 non equivalent pairs of atoms, associated to 61 distances $\{d_j\},j=1\ldots 61$. This defines a $\ND=61$ dimensional "Distance space" $\mathcal{D}$, without translation and rotation degeneracy. 

In both cases, the original dimension of the problem is reduced (from 144 to 100 and 61 respectively) but yet the configuration spaces $\mathcal{X}$ and $\mathcal{D}$ are large and preserve to a large extent the complexity of the original conformations. As our comparisons show, the classification of lipid states is efficient, whether one starts from $\mathcal{X}$ or $\mathcal{D}$. 

Three different algorithms commonly used for Machine Learning classifications were used in this Letter. A short description of these algorithms is given below.

\subsection{Naive Bayes}

Given a configuration space $\mathcal{D} = \{D_j\}, j=1\ldots \ND$, and two classes $s=\{\mathrm{gel},\mathrm{fluid}\}$, the Bayesian approach assumes that it exists a joint probability distribution $P(s,\{D_j\})$, that weighs the respective likelihood of each state $s$ once a configuration $\{D_j\}$ is provided. The Bayesian model makes a decision regarding the state $s$ by comparing the conditional probability densities $P(\mathrm{gel}| \{D_j\})$ and $P(\mathrm{fluid}| \{D_j\})$. To be precise, the Bayesian approach attributes a \textit{fluid} label to a configuration if the sign of 
\beq
\ln \left( \frac{P(\mathrm{fluid}|\{D_j\})}{P(\mathrm{gel}|\{D_j\})}
\right)
\label{eq:BayesianApproachPrinciple}
\eeq
 is positive, and a \textit{gel} label otherwise. 
 
 Each Bayesian approach provides a mathematical  model $P(\{D_j\}|s)$ describing the expected configuration distribution for a given class $s$: \textit{fluid} and \textit{gel}.
There is in principle entire freedom in choosing the model, but the efficiency and the optimization requirements limit  such choices in practice. Training a Bayesian model therefore amounts to finding the most realistic function $P(\{D_j\}|s)$ as far as classifying a given training set of data is concerned. 

The Bayes theorem provides the connection between the conditional probabilities entering in the choice function (\ref{eq:BayesianApproachPrinciple}) and the model:
\beq
P(s|\{D_j\})=\frac{P(\{D_j\}|s) P(s)}{P(\{D_j\})},
\label{eq:BayesTheorem}
\eeq
where $P(s)$  represents the \textit{prior} distribution of $s$, \textit{i.e.} the statistical distribution of $s$ in the absence of any configuration related information, and $P(\{D_j\})= P(\mathrm{gel},\{D_j\} )+P(\mathrm{fluid}, \{D_j\})$ a normalization factor which cancels out in eq~(\ref{eq:BayesianApproachPrinciple}). The Naive Bayes (NB) gaussian model assumes that $P(\{D_j\}|s)$ factorizes as a product of independent gaussian distributions of $D_j$.
\beq
P(\{D_j\}|s)\prod_{j=1}^{\ND}\mathrm{d} D_j = \prod_{j=1}^{\ND} \frac{\mathrm{d} D_j}{\sqrt{2\pi} \sigma_{s,j}}\exp\left({\displaystyle -\frac{(D_j - D_{s,j})^2}{2\sigma_{s,j}^2}} \right)
\label{eq:NaiveBayesGaussianModel}
\eeq
Training the algorithm means finding the best mean value $D_{s,j}$ and standard deviation $\sigma_{s,j}$ for every parameter $D_j$ in the distance space $\mathcal{D}$ and each class $s$. The number of parameters to compute turns out to be equal to twice the dimension of $\mathcal{D}$. In our case, the training set contains an equal number of gel and fluid conformations, and there is no \textit{a priori} bias between classes, meaning that $P(\mathrm{gel}) = P(\mathrm{fluid})=1/2$. Therefore, the number of parameters to determine during training is $2\times 61=122$. 

To sum up, the Naive Bayes approach classifies a lipid conformation by computing a quadratic score function in the conformation space,
\beq
\sum_{j=1}^{\ND} \frac{(D_i-D_{\mathrm{fluid},i})^2}{2\sigma^2_{\mathrm{fluid},i}} -\sum_{j=1}^{\ND} \frac{(D_i-D_{\mathrm{gel},i})}{2\sigma^2_{\mathrm{gel},i}} +\mathrm{Const}
\label{eq:effectiveBayesianDistance}
\eeq
and deciding whether a $\{D_i\}$ lies closer to $\{D_{\mathrm{fluid},i}\}$ or to $\{D_{\mathrm{gel},i}\}$ according to this generalized distance.

\subsection{K-Nearest Neighbors}

The K-Nearest Neighbors (KNN) classification method is based on defining a distance between any arbitrary pairs of objects to discriminate. A natural choice is the Euclidean norm of the feature space, here the $\NX$ dimensional coordinate space $\mathcal{X}$.

The KNN algorithm finds the $K$ nearest neighbors within the training set, of each new configuration to classify. Decision is taken based on majority rule, \textit{i.e.}  the most abundant class found among the $K$ closest neighbors. The optimal $K$ is determined during the training and validation process, and was set equal to 5, a typical value for this method.

\subsection{Support Vector Machines}

Support Vector Machines (SVM) classify data by means of linear separation in high dimensional representation spaces. Denoting $\bm{\phi}$ an arbitrary element in a given representation space $\mathcal{R}$, the binary decision is given by the sign of the affine expression $\bm{w}\cdot \bm{\phi}+b$, with $b$ a numerical constant and $\bm{w}$ the hyperplane of separation normal vector. In a few favorable cases, it is possible to use directly the data definition space as representation space. However, in many practical situations, efficient classification can only be achieved by mapping the data (\textit{e.g.} $\mathcal{X}$ or $\mathcal{D}$) onto a larger representation space, namely $\bm{x}\mapsto \bm{\phi}(\bm{x})$. Training a SVM  corresponds to choosing a suitable representation space $\mathcal{R}$, and finding the optimal $b$ and $\bm{w}$. As shown in \cite{boser_vapnik_1992,cortes_vapnik_1995}, given a training set $\{\bm{x}_i\}, i=1\ldots n$, the optimal $\bm{w}$ can always be expressed as a linear combination $\bm{w} = \sum_i \alpha_i y_i \bm{\phi}(\bm{x}_i)$ with either positive or vanishing $\alpha_i$ coefficients, and $y_i= \pm 1$, depending on the class (\textit{fluid} 1, \textit{gel} -1) of the corresponding vector data $\bm{x}_i$.

The subset of vectors $\{\bm{x}_i\}$ participating in the definition of $\bm{w}$ with non vanishing coefficients $\alpha_i>0$ forms the so-called \textit{support vectors}. Denoting $\mathcal{J}$ the sequence of indices  of support vectors, and $\mathcal{K}(\bm{x},\bm{x}')$ the product $\bm{\phi}(\bm{x})\cdot\bm{\phi}(\bm{x}')$ in $\mathcal{R}$, called kernel function, the SVM decision function for any vector data $\bm{x}$ reads:
\begin {equation}
\mathrm{sign}\left(b+\sum_{j\in\mathcal{J}} \alpha_j y_j \mathcal{K}(\bm{x}_j,\bm{x})\right).
\label{eq:SVMDecisionFunction}
\end{equation}
The SVM training optimization problem can therefore be formulated without any explicit reference to the representation space $\mathcal{R}$, nor the mapping $\bm{\phi}(x)$. It only requires an explicit positive kernel function $\mathcal{K}(\bm{x},\bm{x}')$. As explained in \cite{boser_vapnik_1992,cortes_vapnik_1995}, there are efficient quadratic optimization algorithms for determining the support vectors $\bm{x}_j$, the non-vanishing coefficients $\alpha_j$  and the shift constant $b$. 

In this study, we used the standard radial basis kernel function 
\begin{equation}
\mathcal{K}(\bm{x},\bm{x}') = \exp\left(-\gamma ||\bm{x}-\bm{x}'||^2\right),   
\end{equation}
with a default value $\gamma$ equal to the inverse of the dimension of the configuration space. This choice assumes that each component of $\bm{x}$ is of order~1. When using the coordinate space $\mathcal{X}$, the trained SVM  ends up using 297 non vanishing support vectors and coefficients $\alpha_j$, $\gamma =1/100$, $b=-0.416$ (167~\textit{fluid}, 130~\textit{gel} support vectors). When considering the distance space $\mathcal{D}$, the trained SVM used 161 non vanishing coefficients $\alpha_j$, $\gamma = 1/61$ and $b = 0.134$ (100~\textit{fluid}, 61~\textit{gel} support vectors).

\section{Training}

The Machine Learning analysis performed in this Letter were conducted using the \texttt{Scikit-Learn} module for Python 3~\cite{pedregosa_duchesnay_2011}. An unbiased selection of lipid conformations extracted from a trajectory at low temperature (288~K) was part of the training set, with a label \textit{gel}. Similarly, an unbiased selection of conformations from a trajectory at high temperature (358~K) was added to the training set with a label \textit{fluid}. The number of gel and fluid conformations were in equal number in the training set. As customary, 20~\% of conformations were removed from the training sets, and used for verification and scoring purposes.  The training set therefore consists of a sequence of 370 vectors (elements of the  configuration spaces $\mathcal{X}$ or $\mathcal{D}$), used for building the prediction model. 

%%%%%%%%%%%%%%%%%%%%%%%%%%%%%%%%%%%%%%%%%%%%%%%%%%%%%%%%%%%%%%%
%\section{Combination of classifications}
\section{Asserting the predictive capacity of each model}

Once defined the training set, with properly labelled gel and fluid states, 20\% of the lipids in each phase were taken apart for forming a validation set. After training the models (\textit{i.e.} optimizing the parameters with respect to the 80\% remaining conformations) a prediction score was separately calculated for the \textit{gel} and \textit{fluid} conformations in the validation set. The overall procedure was repeated 10 times, each time with the same training set, but independently drawn validation subsets.

%%%%%%%%%%%%%%%%%%%%%%%%%%%%%%%%%%%%%%%%%%%%%%%%%%%%%%%%%
\begin{figure}
\centering{
\begin{tabular}{cc}
\subfloat{\includegraphics[scale=.45]{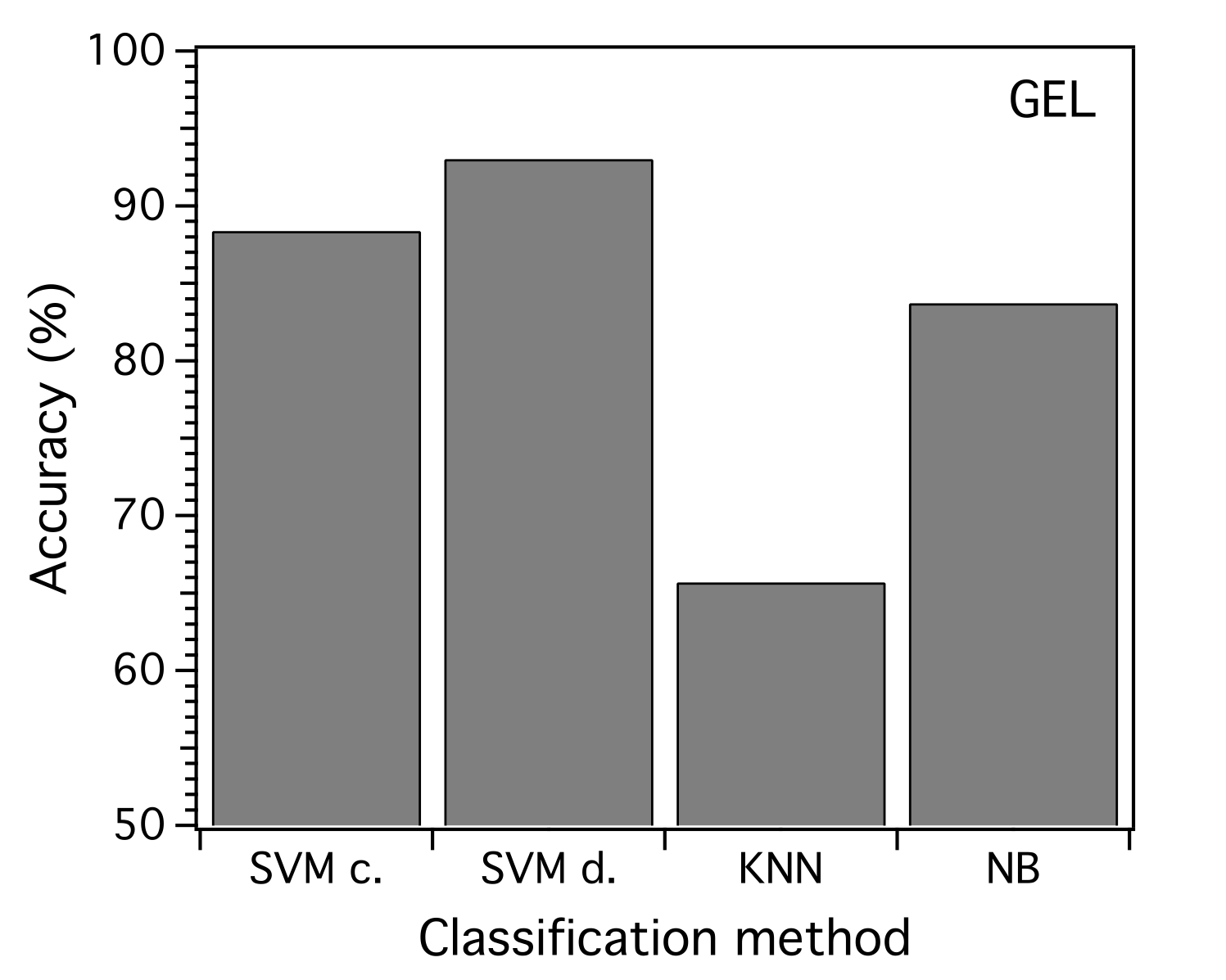}} &
\subfloat{\includegraphics[scale=.45]{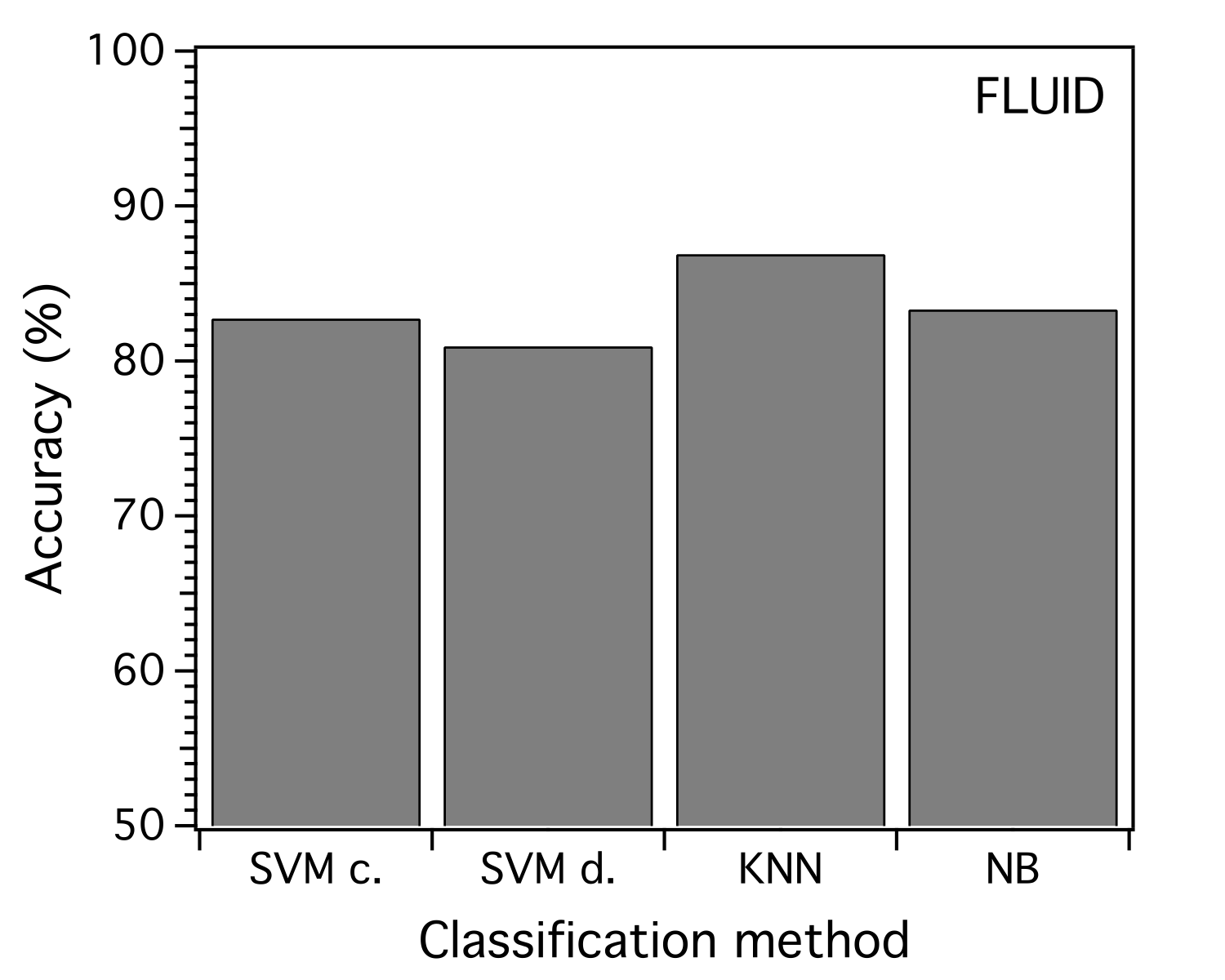}}
\end{tabular}
\caption{Prediction scores of the Machine Learning classification methods for lipids in the gel phase (Left) and in the fluid phase (Right). Besides the Naive Bayes method (NB), all methods have an important asymmetry in their accuracy between each phase. We call these asymmetries the "expertises" of the methods.}
\label{sfig_expertises}
}
\end{figure}
%%%%%%%%%%%%%%%%%%%%%%%%%%%%%%%%%%%%%%%%%%%%%%%%%%%%%

The resulting scores shown in Fig.~\ref{sfig_expertises} shows that the NB predictive capacity is independent from the phase of the lipid considered. KNN performs better for fluids than for gels. At the opposite, ''coordinates'' $\mathcal{X}$-SVM and ''distances'' $\mathcal{D}$-SVM seems to perform better in the gel phase. 

We therefore decided to combine the predictions of the above models, retaining those who perform best in each phase. The following decision chain was implemented:

\begin{enumerate}
    \item If the 4 models agree on the same prediction for the phase, this prediction is retained;
    \item if the $\mathcal{D}$-SVM algorithm predicts a gel phase, the lipid configuration is assumed to be \textit{gel};
    \item if the $\mathcal{X}$-SVM algorithm predicts a fluid phase, the lipid configuration is assumed to be \textit{fluid};
    \item if the NB algorithm predicts a gel phase, the lipid configuration is assumed to be \textit{gel}; 
    \item if none of the above conditions have been met, the KNN algorithm makes the final decision on the configuration classification.
\end{enumerate}

\section{Comparison between Machine Learning decisions and structural characterizations of the lipid configurations}

\begin{figure}
\centering{
\begin{tabular}{cc}
\subfloat{\includegraphics[scale=.49]{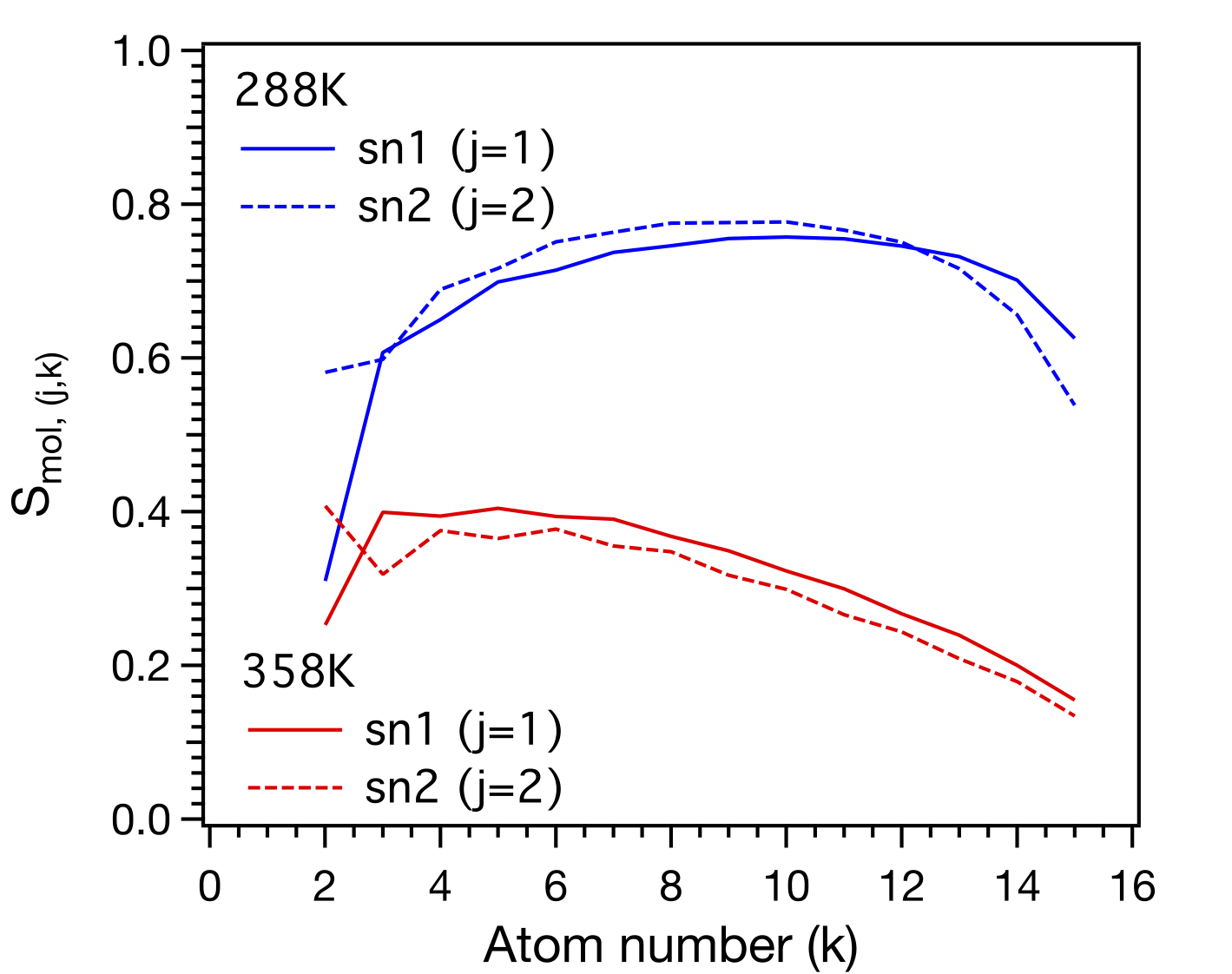}} &
\subfloat{\includegraphics[scale=.48]{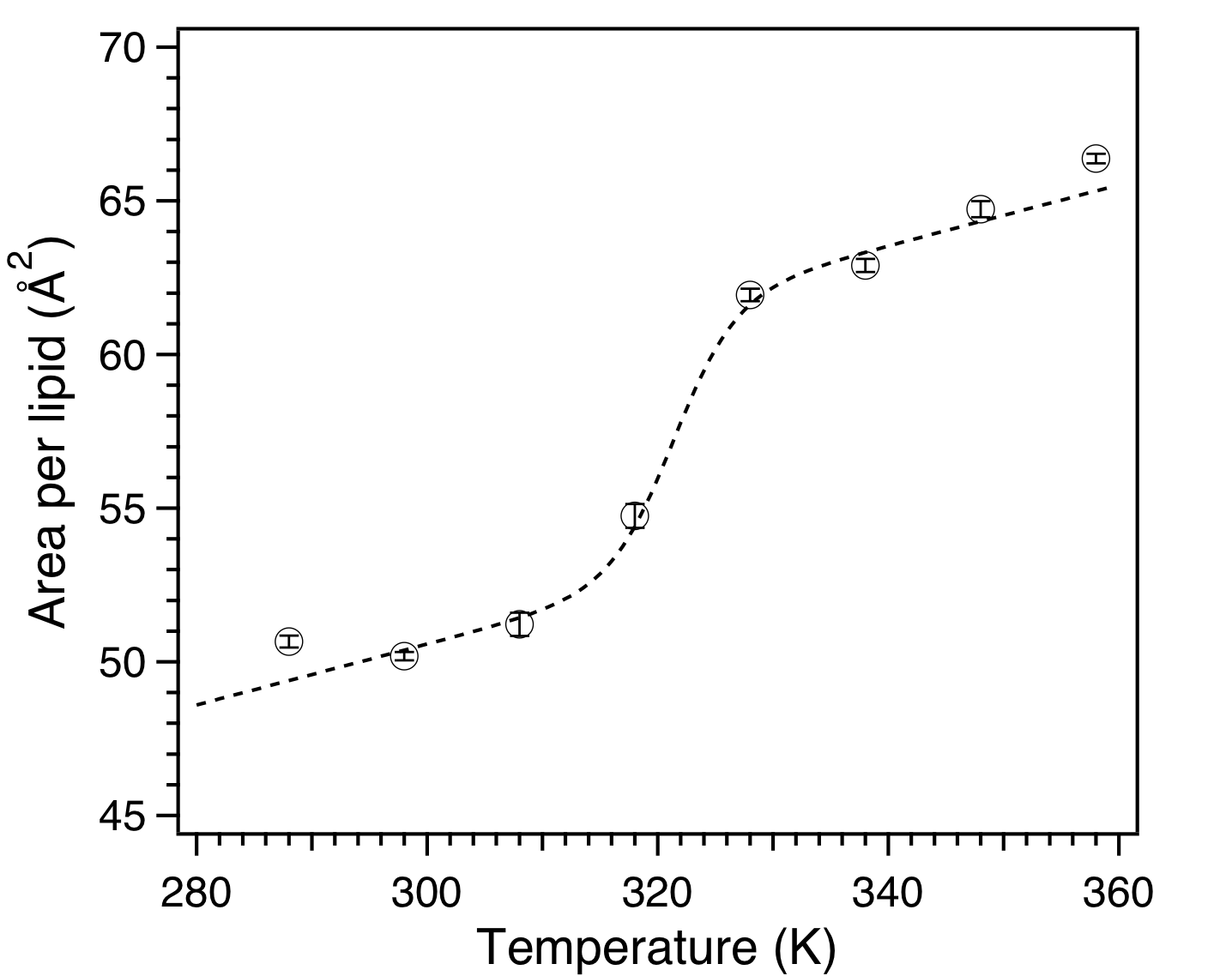}}
\end{tabular}
\caption{Confirmation of the thermodynamic phase of the bilayer using two common structural parameters: (a) the order parameter $S_{\mathrm{mol},(j,k)}$ with $k$ the atom number, $j=1$ (tail sn1) or  $j=2$ (tail sn2), and (b) the area per lipid $A_l$. (Left) the average order parameter is shown as a function of the carbon atom index along the chain (from glycerol to terminal end), for each \textit{sn1} and \textit{sn2} chain. (Right) the phase transition can be clearly seen in the evolution of the area per lipid as a function of temperature. A sigmoid fit points to a transition temperature $T_m$ equal to 321~K in our system.}
\label{sfig_verification}
}
\end{figure}

Machine Learning predictions were compared to two typical lipid structural properties: the carbon carbon (CC) order parameter $S_{\mathrm{mol}}$ along the chains and the area per lipid $A$ in the 2d Voronoi tessellation of the lipid projected centers of mass. The corresponding results are presented in Fig.~\ref{sfig_verification}(a) and (b).

%%%%%%%%%%%%%%%%%%%%%%%%%%%%%%%%%%%%%%%%%%%%%%%%%%%%%
%%%%%%%%%%%%%%%%%%%%%%%%%%%%%%%%%%%%%%%%%%%%%%%%%%%

The order parameter curves (Fig.~\ref{sfig_verification}(a)) clearly discriminate among the low temperature (288~K) and high temperature (358~K) fluid phases, in agreement with published results on these systems~\cite{klauda_pastor_2010}.

The order parameter of the atoms in the lipid tails at low (288~K) and high (358~K) temperature is characteristic from membranes in the gel and fluid phases respectively \cite{Cevc_Marsh_PhospholipidBilayers}. The phase transition can be clearly seen in Fig.~\ref{sfig_verification}(b) as a significant variation in the evolution of the area per lipid $A_l$ around 321~K. 

 Experimental structural values are available at 323~K~\cite{nagle_suter_1996, nagle_tristram_2000, kucerka_nagle_2006, kucerka_katsaras_2008}. Nagle \textit{et al.} obtained for DPPC an area per lipid equals to 64 $\pm$ 1~\AA$^2$ significantly close to the value \textit{circa} 60~\AA$^2$ we obtained in our simulations. Using the average DPPC bilayer thickness reported by Nagle \textit{et al.}, we could estimate an experimental volume per lipid of 1220 $\pm$ 50~\AA$^3$, which agrees fairly with our Voronoi value of 1300~\AA$^3$

%%%%%%%%%%%%%%%%%%%%%%%%%%%%%%%%%%%%%%%%%%%%%%%%%%%%%%%
\section{Naive classifications}

\begin{figure}
\centering{
\begin{tabular}{cc}
\subfloat{\includegraphics[scale=.48]{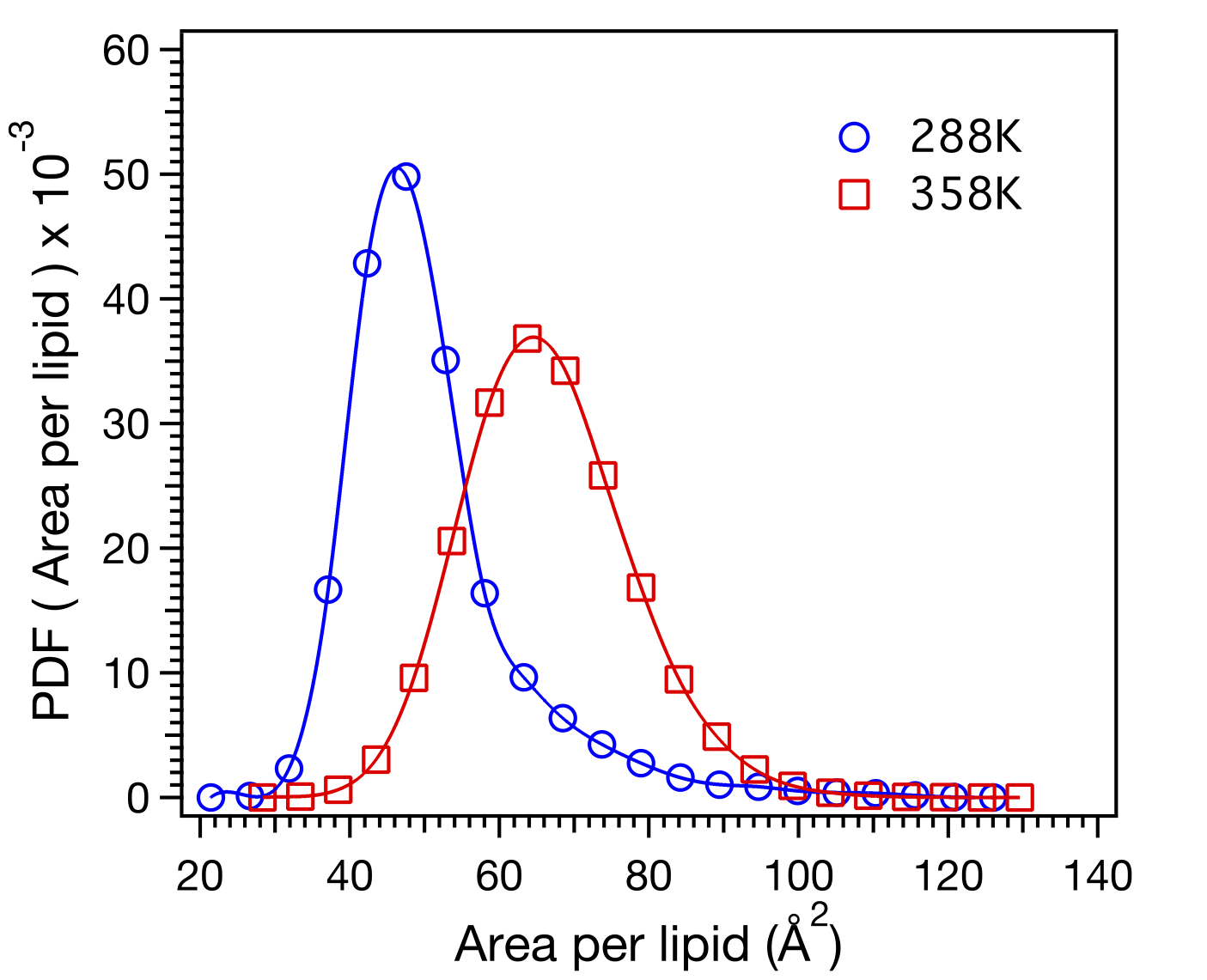}} &
\subfloat{\includegraphics[scale=.45]{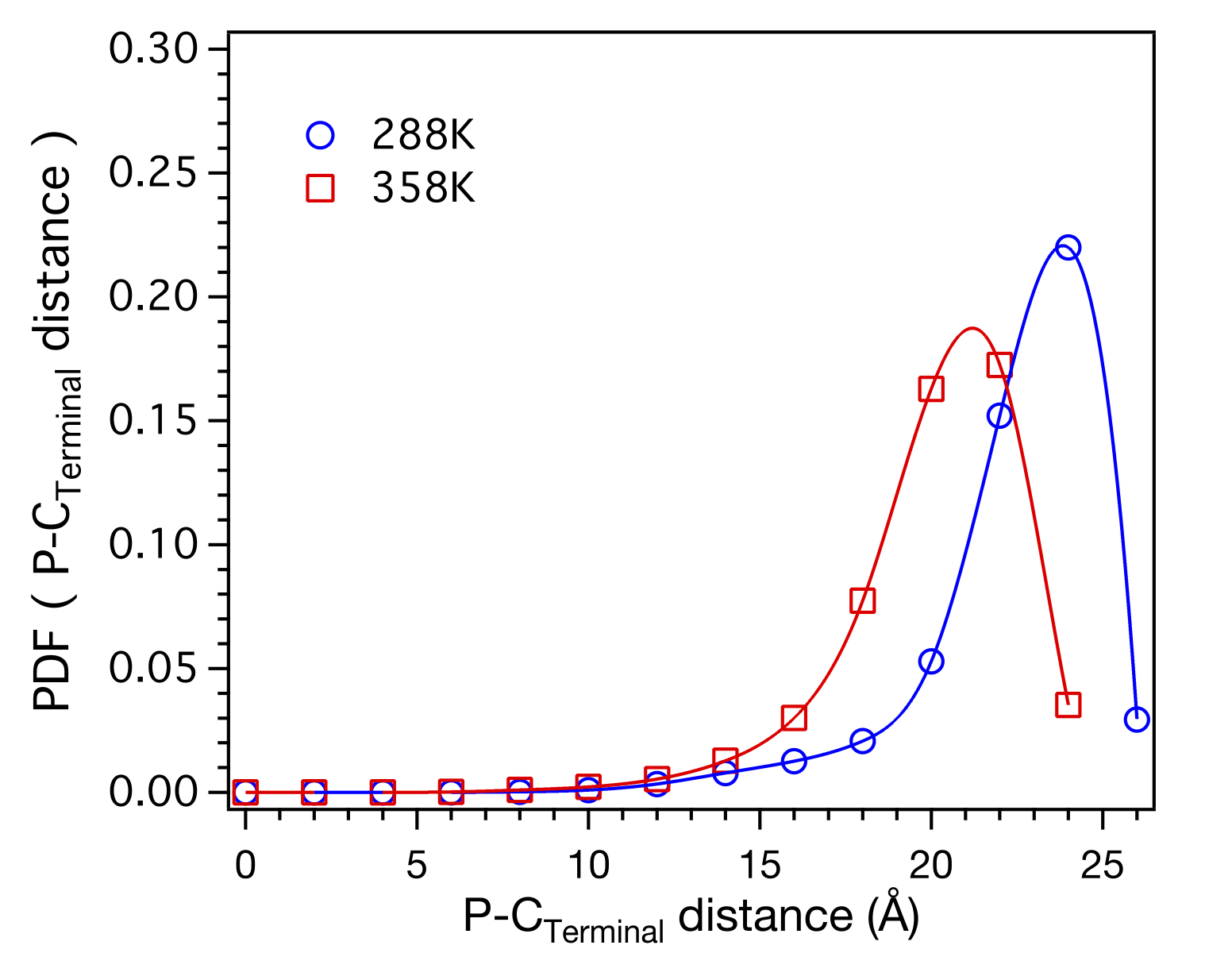}}
\end{tabular}
\caption{Distributions of the area per lipid (Left) and the average elongation between phosphorus and \textit{sn1} terminal carbon atoms (Right) from lipids conformations at 288 and 358~K. }
\label{sfig_naive_discrimination}
}
\end{figure}

The distributions of the areas per lipid and molecular elongations at low and high temperatures are shown in Fig.~\ref{sfig_naive_discrimination}. Using a naive classification scheme based on a single threshold value for either of the two previous scalar parameters would at best result in a prediction  accuracy of respectively 69\% and 67\%.

\begin{figure}
\centering{
\includegraphics[scale=.48]{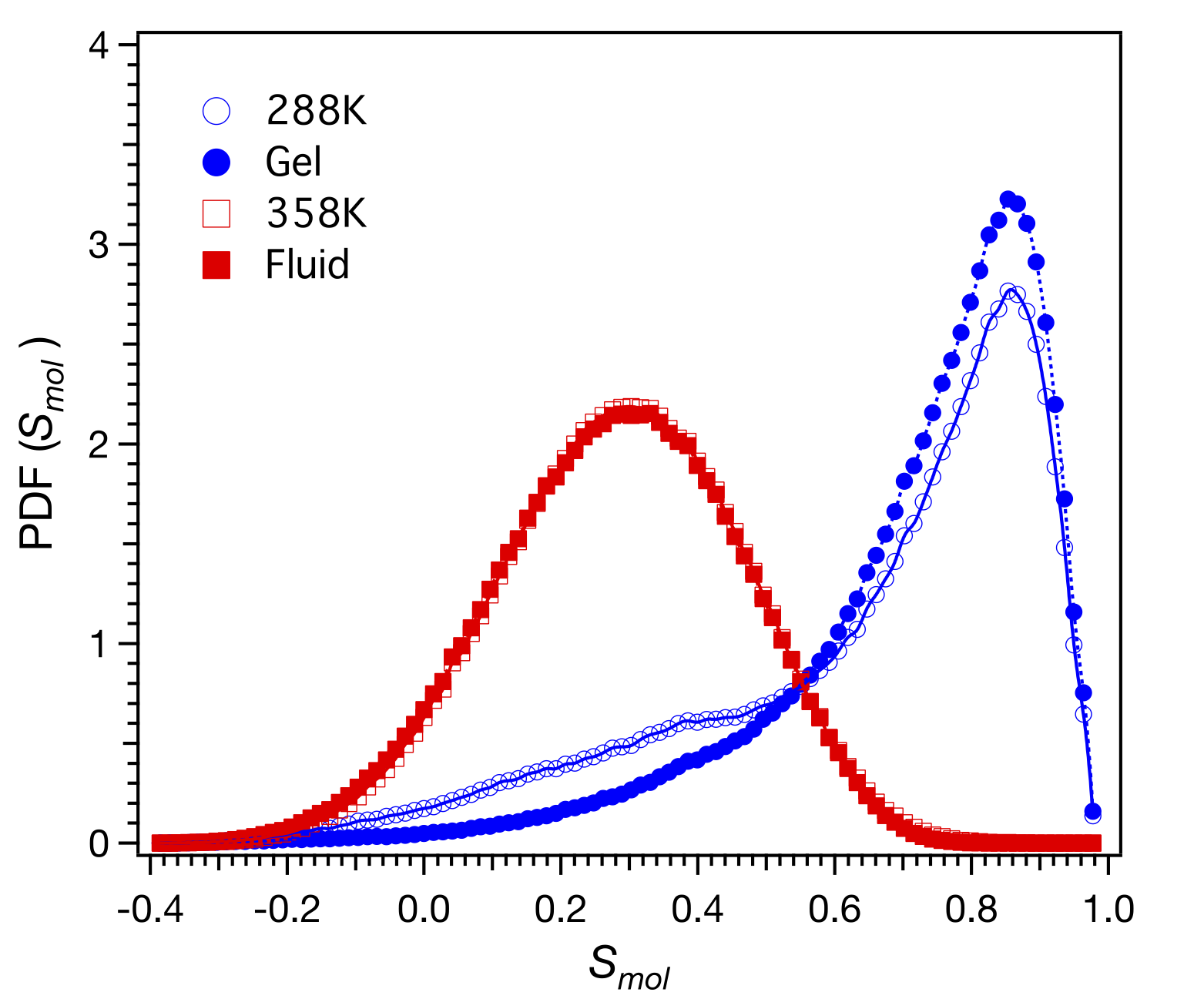}
\caption{Histograms of the molecular order parameter $S_{\mathrm{mol}}$ of lipids sorted by temperature (blue circles) and state ML classification (red squares).}
\label{sfig_discrimination_postml}
}
\end{figure}

Fig.~\ref{sfig_discrimination_postml} compares the histogram of molecular order parameters $S_{\mathrm{mol}}$ as a function of the temperature of the lipid bilayer from which the configurations are extracted (288~K or 358~K), and as a function of the result of the Machine Learning classification procedure. The difference between the distribution at 288~K and the distribution in the \textit{a posteriori} gel state ensemble indicates that a small fraction of lipids in the fluid state are already present at 288~K.

%%%%%%%%%%%%%%%%%%%%%%%%%%%%%%%%%%%%%%%%%%%%%%%%%%%%%%%

%%%%%%%%%%%%%%%%%%%%%%%%%%%%%%%%%%%%%%%%%%%%%%%%%%%%%%%%


\begin{thebibliography}{10}

\bibitem{Mouritsen_MatterOfFat}
O. G. Mouritsen.
\newblock {\em Life-as a matter of fat: the emerging science of lipidomics}.
\newblock Springer, 2005.

\bibitem{Cevc_Marsh_PhospholipidBilayers}
G. Cevc and D. Marsh.
\newblock {\em Phospholipid Bilayers. Physical Principles and Models}.
\newblock John Wiley \& Sons, New-York, 1987.

\bibitem{Dimova_Marques_TheGiantVesicleBook}
R. Dimova and C.M. Marques
\newblock{\em The Giant Vesicle Book}.
\newblock CRC Press, Taylor \& Francis, 2019.

\bibitem{1976_Mabrey_Sturtevant}
S. Mabrey and J.~M. Sturtevant.
\newblock Investigation of phase transitions in lipids and lipid mixtures by high sensitivity differential scanning calorimetry.
\newblock {\em Proceding of the Natural Academy of Sciences USA},
  73(11):3862--3866, 1976.

\bibitem{Heimburg_BiophysicsMembrane}
T. Heimburg.
\newblock {\em Thermal Biophysics of Membranes}.
\newblock Wiley-VCH, 2007.

\bibitem{Marsh_HandbookLipidBilayers2}
D. Marsh.
\newblock {\em Handbook of Lipid Bilayers}.
\newblock CRC Press, Boca Raton, 2nd edition, 2013.

\bibitem{2012_Armstrong_Rheinstadter}
C.~L. Armstrong, M.~A. Barrett, L.~Toppozini, N.~Kucerka, Z.~Yamani,
  J.~Katsaras, G.~Fragneto, and M.~C. Rheinstadter.
\newblock Co-existence of gel and fluid lipid domains in single-component
  phospholipid membranes.
\newblock {\em Soft Matter}, 8:4687--4694, 2012.

\bibitem{1973_Nagle}
J.~F. Nagle.
\newblock The theory of biomembrane phase transitions.
\newblock {\em Journal of Chemical Physics}, 58(1):252, 1973.

\bibitem{1974_Marsh}
D.~Marsh.
\newblock Statistical mechanics of the fluidity of phospholipid bilayers and
  membranes.
\newblock {\em Journal of Membrane Biology}, 18:145--162, 1974.

\bibitem{1974_Marcelja_2}
S.~Marcelj\`{a}.
\newblock Chain ordering in liquid crystals. ii structure of bilayer membranes.
\newblock {\em Biochim. Biophys. Acta}, 367:165--176, 1974.

\bibitem{1979_Pink_Chapman}
D.~A. Pink and D. Chapman.
\newblock {Protein-lipid interactions in bilayer membranes: A lattice model}.
\newblock {\em Proceedings of the National Academy of Sciences of the United
  States of America}, 76(4):1542--1546, 1979.

\bibitem{1978_Doniach}
S.~Doniach.
\newblock Thermodynamic fluctuations in phospholipid bilayers.
\newblock {\em The Journal of Chemical Physics}, 68(11):4912--4916, 1978.

\bibitem{1983_Sugar_Monticelli}
I.~P. Sug\`{a}r and G. Monticelli.
\newblock Landau theory of two-component phospholipid bilayers. i
  phosphatidylcholine/phosphatidylethanolamine mixtures.
\newblock {\em Biophysical Chemistry}, 18:281--289, 1983.

\bibitem{1996_Heimburg_Biltonen}
T.~Heimburg and R.~L. Biltonen.
\newblock A monte carlo simulation study of protein-induced heat capacity
  changes and lipid-induced protein clustering.
\newblock {\em Biophysical Journal}, 70(1):84 -- 96, 1996.

\bibitem{1996_Jerala_Biltonen}
R. Jerala, P.~F.~F. Almeida, and R. Biltonen.
\newblock Simulation of the gel-fluid transition in a membrane composed of
  lipids with two connected acyl chains: Application of a dimer-move step.
\newblock {\em Biophysical Journal}, 71:609--615, 1996.

\bibitem{1999_Sugar_Biltonen}
I.~P. Sug\`{a}r, T.~E. Thompson, and R.~L. Biltonen.
\newblock Monte-carlo simulation of two-component bilayers:dmpc/dspc mixtures.
\newblock {\em Biophysical Journal}, 76:2099--2110, 1999.

\bibitem{2001_Ivanova_Heimburg}
V.~P. Ivanova and T. Heimburg.
\newblock Histogram methods to obtain heat capacities in lipid monolayers,
  curved bilayers and membrane containing peptides.
\newblock {\em Physical Review E}, 63:041914, 2001.

\bibitem{2011_Wolff_Thalmann}
J.~Wolff, C.~M. Marques, and F.~Thalmann.
\newblock Thermodynamic approach to phase coexistence in ternary
  phospholipid-cholesterol mixtures.
\newblock {\em Phys. Rev. Lett.}, 106(12):128104, 2011.

\bibitem{2018_Morandi_Marques}
M.~I. Morandi, M. Sommer, M. Kluzek, F. Thalmann, A.~P.
  Schroder, and C.~M. Marques.
\newblock Dppc bilayers in solutions of high sucrose content.
\newblock {\em Biophysical Journal}, 114(9):2165 -- 2173, 2018.

\bibitem{2015_Cubuk_Liu}
E.~D. Cubuk, S.~S. Schoenholz, J.~M. Rieser, B.~D. Malone, J.~Rottler, D.~J.
  Durian, E.~Kaxiras, and A.~J. Liu.
\newblock Identifying structural flow defects in disordered solids using
  machine-learning methods.
\newblock {\em Phys. Rev. Lett.}, 114:108001, 2015.

\bibitem{2017_Carrasquilla_Melko}
J. Carrasquilla and R.~G. Melko.
\newblock Machine learning phases of matter.
\newblock {\em Nature Physics}, 13:431, 2017.

\bibitem{2019_Le_Tran}
T.~C. Le and N. Tran.
\newblock Using machine learning to predict the self-assembled nanostructures
  of monoolein and phytantriol as a function of temperature and fatty acid
  additives for effective lipid-based delivery systems.
\newblock {\em ACS Applied Nano Materials}, 2(3):1637--1647, 2019.

\bibitem{2010_Klauda_Pastor}
J.~B. Klauda, R.~M. Venable, J.~A. Freites, J.~W.
  O’Connor, Douglas~J. Tobias, C. Mondragon-Ramirez, I. Vorobyov,
  A.~D. MacKerell, and R.~W. Pastor.
\newblock Update of the charmm all-atom additive force field for lipids:
  Validation on six lipid types.
\newblock {\em The Journal of Physical Chemistry B}, 114(23):7830--7843, 2010.

\bibitem{best_mackerell_2012}
R.~B. Best, X.~Zhu, J.~Shim, P.~E.~M. Lopes, J.~Mittal, M.~Feig, and A.~D.
  MacKerell~Jr.
\newblock Optimization of the additive charmm all-atom protein force field
  targeting improved sampling of the backbone phi, psi and side-chain khi1 and
  khi2 dihedral angles.
\newblock {\em Journal of Chemical Theory and Computation}, 8(9):3257--3273,
  2012.

\bibitem{2005_deVries_Marrink}
A.~H. de~Vries, S. Yefimov, A.~E. Mark, and S.~J. Marrink.
\newblock Molecular structure of the lecithin ripple phase.
\newblock {\em Proceedings of the National Academy of Sciences},
  102(15):5392--5396, 2005.

\bibitem{2018_Khakbaz_Klauda}
P. Khakbaz and J.~B. Klauda.
\newblock Investigation of phase transitions of saturated phosphocholine lipid
  bilayers via molecular dynamics simulations.
\newblock {\em Biochimica et Biophysica Acta (BBA) - Biomembranes},
  1860(8):1489 -- 1501, 2018.

\bibitem{pedregosa_duchesnay_2011}
F.~Pedregosa, G.~Varoquaux, A.~Gramfort, V.~Michel, B.~Thirion, O.~Grisel,
  M.~Blondel, P.~Prettenhofer, R.~Weiss, V.~Dubourg, J.~Vanderplas, A.~Passos,
  D.~Cournapeau, M.~Brucher, M.~Perrot, and E.~Duchesnay.
\newblock Scikit-learn: Machine learning in {P}ython.
\newblock {\em Journal of Machine Learning Research}, 12:2825--2830, 2011.

\bibitem{1963_Glauber}
R.~J. Glauber.
\newblock Time‐dependent statistics of the ising model.
\newblock {\em Journal of Mathematical Physics}, 4(2):294--307, 1963.

\bibitem{1988_Ipsen_Mouritsen}
J.~H. Ipsen and O.~G. Mouritsen.
\newblock Modelling the phase equilibria in two-components membranes of
  phospholipids with different acyl-chain lengths.
\newblock {\em Biochim. Biophys. Acta}, 944:121--134, 1988.

\bibitem{1987_Ipsen_Zuckermann}
J.~H. Ipsen, G.~Karlstr{\"o}m, O.~G. Mouritsen, H.~Wennerstr{\"o}m, and M.J.
  Zuckermann.
\newblock Phase equilibria in phosphatidylcholine-cholesterol system.
\newblock {\em Biochim. Biophys. Acta}, 905:162--172, 1987.

\bibitem{2014_Ingolfsson_Marrink}
H.~I. Ing\`{o}lfsson, M.~N. Melo, F.~J. van Eerden, C. Arnarez, C.~A. Lopez, T.~A. Wassenaar, X. Periole, A.~H. de~Vries, D.~P. Tieleman, and S.~J. Marrink.
\newblock Lipid organization of the plasma membrane.
\newblock {\em Journal of the American Chemical Society}, 136(41):14554--14559,
  2014.

\bibitem{2014_Franova_Ollila}
M.~D. Fra\v{n}ov\`{a}, I. Vattulainen, and O.~H.~S.  Ollila.
\newblock Can pyrene probes be used to measure lateral pressure profiles of
  lipid membranes? perspective through atomistic simulations.
\newblock {\em Biochimica et Biophysica Acta (BBA) - Biomembranes},
  1838(5):1406 -- 1411, 2014.

\bibitem{1991_Crowe_Crowe}
L.~M. Crowe and J.~H. Crowe.
\newblock Solution effects on the thermotropic phase transition of unilamellar
  liposomes.
\newblock {\em Biochimica et Biophysica Acta (BBA) - Biomembranes}, 1064(2):267 -- 274, 1991.

\bibitem{2014_Rossi_Monticelli}
G. Rossi, J. Barnoud, and L. Monticelli.
\newblock Polystyrene nanoparticles perturb lipid membranes.
\newblock {\em The Journal of Physical Chemistry Letters}, 5(1):241--246, 2014.

\bibitem{2014_Rossi_Monticelli_2}
G. Rossi and L. Monticelli.
\newblock Modeling the effect of nano-sized polymer particles on the properties of lipid membranes.
\newblock {\em Journal of Physics: Condensed Matter}, 26(50):503101, 2014.

\bibitem{1997_Gil_Mouritsen}
T.~Gil, M.~C.~Sabra, J.~H.~Ipsen, and O.~G.~Mouritsen.
\newblock Wetting and capillary condensation as means of protein organization
  in membranes.
\newblock {\em Biophysical Journal}, 73(4):1728 -- 1741, 1997.

\bibitem{1998_Gil_Zuckermann}
T. Gil, J.~H. Ipsen, O.~G. Mouritsen, M.~C. Sabra, M.~M. Sperotto, and M.~J. Zuckermann.
\newblock Theoretical analysis of protein organization in lipid membranes.
\newblock {\em Biochimica et Biophysica Acta (BBA) - Reviews on Biomembranes},
  1376(3):245 -- 266, 1998.

\end{thebibliography}

\begin{thebibliography}{10}

\bibitem{berendsen_vandrunen_1995}
H~Berendsen, D~van~der Spoel, and R~van Drunen.
\newblock Gromacs: A message-passing parallel molecular dynamics
  implementation.
\newblock {\em Computer Physics Communications}, 1995.

\bibitem{abraham_lindahl_2015}
M~J Abraham, T~Murtola, R~Schulz, S~P\'all, J~C Smith, B~Hess, and E~Lindahl.
\newblock Gromacs: High performance molecular simulations through multi-level
  parallelism from laptops to supercomputers.
\newblock {\em SoftwareX}, 2015.

\bibitem{best_mackerell_2012}
R.~B. Best, X.~Zhu, J.~Shim, P.~E.~M. Lopes, J.~Mittal, M.~Feig, and A.~D.
  MacKerell~Jr.
\newblock Optimization of the additive charmm all-atom protein force field
  targeting improved sampling of the backbone phi, psi and side-chain khi1 and
  khi2 dihedral angles.
\newblock {\em Journal of Chemical Theory and Computation}, 8(9):3257--3273,
  2012.

\bibitem{jo_im_2008}
S.~Jo, T.~Kim, V.~G. Iyer, and W.~Im.
\newblock Charmm-gui: A web-based graphical user interface for charmm.
\newblock {\em Journal of Computational Chemistry}, 29:1859--1865, 2008.

\bibitem{wu_im_2014}
E.~L. Wu, X.~Cheng, S.~Jo, H.~Rui, H.~K. Song, E.~M. Davila-Contreras, Y.~Qi,
  J.~Lee, V.~Monje-Galvan, R.~M. Venable, J.~B. Klauda, and W.~Im.
\newblock Charmm-gui membrane builder toward realistic biological membrane
  simulations.
\newblock {\em Journal of Chemical Theory and Computation}, 35:1997--2004,
  2014.

\bibitem{jo_im_2009}
S.~Jo, J.~B. Lim, J.~B. Klauda, and W.~Im.
\newblock Charmm-gui membrane builder for mixed bilayers and its application to
  yeast membranes.
\newblock {\em Biophysical Journal}, 97:50--58, 2009.

\bibitem{jo_im_2007}
S.~Jo, T.~Kim, and W.~Im.
\newblock Automated builder and database of protein/membrane complexes for
  molecular dynamics simulations.
\newblock {\em PLoS ONE}, 2(9):880, 2007.

\bibitem{brooks_karplus_2009}
B.~R. Brooks, C.~L. Brooks~III, A.~D. MacKerell~Jr, L.~Nilsson, R.~J. Petrella,
  B.~Roux, Y.~Won, G.~Archontis, C.~Bartels, S.~Boresch, A.~Caflisch, L.~Caves,
  Q.~Cui, A.~R. Dinner, M.~Feig, S.~Fischer, J.~Gao, M.~Hodoscek, W.~Im,
  K.~Kuczera, T.~Lazaridis, J.~Ma, V.~Ovchinnikov, E.~Paci, R.~W. Pastor, C.~B.
  Post, J.~Z. Pu, M.~Schaefer, B.~Tidor, R.~M. Venable, H.~L. Woodcock, X.~Wu,
  W.~Yang, D.~M. York, and M.~Karplus.
\newblock Charmm: The biomolecular simulation program.
\newblock {\em Journal of Computational Chemistry}, 30:1545--1614, 2009.

\bibitem{lee_im_2016}
J.~Lee, X.~Cheng, J.~M. Swails, M.~S. Yeom, P.~K. Eastman, J.~A. Lemkul,
  S.~Wei, J.~Buckner, J.~C. Jeong, Y.~Qi, S.~Jo, V.~S. Pande, D.~A. Case, C.~L.
  Brooks~III, A.~D. MacKerell~Jr, J.~B. Klauda, and W.~Im.
\newblock Charmm-gui input generator for namd, gromacs, amber, openmm, and
  charmm/openmm simulations using the charmm36 additive force field.
\newblock {\em Journal of Chemical Theory and Computation}, 12(1):405--413,
  2016.

\bibitem{hockney_eastwood_1974}
R.~W. Hockney, S.~P. Goel, and J.~W. Eastwood.
\newblock Quiet high-resolution computer models of a plasma.
\newblock {\em Journal of Computational Physics}, 14(2):148--158, 1974.

\bibitem{nose_1984}
S.~Nose.
\newblock A molecular dynamics method for simulations in the canonical
  ensemble.
\newblock {\em Molecular Physics}, 52(2):255--268, 1984.

\bibitem{hoover_1985}
W.~G. Hoover.
\newblock Canonical dynamics: Equilibrium phase-space distributions.
\newblock {\em Physical Review A}, 31:1695, 1985.

\bibitem{nose_klein_1983}
S.~Nose and M.~L. Klein.
\newblock Constant pressure molecular dynamics for molecular systems.
\newblock {\em Molecular Physics}, 50:1055--1076, 1983.

\bibitem{parrinello_rahman_1998}
M.~Parinello and A.~Rahman.
\newblock Polymorphic transitions in single crystals: A new molecular dynamics
  method.
\newblock {\em Journal of Applied Physics}, 52:7182, 1998.

\bibitem{gowers_beckstein_2016}
R~J Gowers, M~Linke, J~Barnoud, T~J~E Reddy, M~N Melo, S~L Seyler, D~L Dotson,
  J~Domanski, S~Buchoux, I~M Kenney, and O~Beckstein.
\newblock Mdanalysis: A python package for the rapid analysis of molecular
  dynamics simulations.
\newblock {\em Proceedings of the 15th Python in Science Conference}, pages
  98--105, 2016.

\bibitem{michaud_beckstein_2011}
N~Michaud‐Agrawal, E~J Denning, T~B Woolf, and O~Beckstein.
\newblock Mdanalysis: A toolkit for the analysis of molecular dynamics
  simulations.
\newblock {\em Journal of Computational Chemistry}, 32(10):2319--2327, 2011.

\bibitem{rycroft_2009}
C~R Rycroft.
\newblock Voro++: A three-dimensional voronoi cell library in c++.
\newblock {\em Chaos}, 19:041111, 2009.

\bibitem{boser_vapnik_1992}
Bernhard~E Boser, Isabelle~M Guyon, and Vladimir~N. Vapnik.
\newblock A training algorithm for optimal margin classifiers.
\newblock In {\em COLT '92 Proceedings of the fifth annual workshop on
  Computational learning theory}, pages 144--152, 1992.

\bibitem{cortes_vapnik_1995}
Corinna Cortes and Vladimir Vapnik.
\newblock Support-vector networks.
\newblock {\em Machine Learning}, 20:273--297, 1995.

\bibitem{pedregosa_duchesnay_2011}
F.~Pedregosa, G.~Varoquaux, A.~Gramfort, V.~Michel, B.~Thirion, O.~Grisel,
  M.~Blondel, P.~Prettenhofer, R.~Weiss, V.~Dubourg, J.~Vanderplas, A.~Passos,
  D.~Cournapeau, M.~Brucher, M.~Perrot, and E.~Duchesnay.
\newblock Scikit-learn: Machine learning in {P}ython.
\newblock {\em Journal of Machine Learning Research}, 12:2825--2830, 2011.

\bibitem{klauda_pastor_2010}
Jeffery~B. Klauda, Richard~M. Venable, J.~Alfredo Freites, Joseph~W.
  O’Connor, Douglas~J. Tobias, Carlos Mondragon-Ramirez, Igor Vorobyov,
  Alexander~D. MacKerell, and Richard~W. Pastor.
\newblock Update of the charmm all-atom additive force field for lipids:
  Validation on six lipid types.
\newblock {\em The Journal of Physical Chemistry B}, 114(23):7830--7843, 2010.
\newblock PMID: 20496934.

\bibitem{Cevc_Marsh_PhospholipidBilayers}
Gregor Cevc and Derek Marsh.
\newblock {\em Phospholipid Bilayers. Physical Principles and Models}.
\newblock John Wiley \& Sons, New-York, 1987.

\bibitem{nagle_suter_1996}
J~F Nagle, R~Zhang, S~Tristram-Nagle, W~Sun, H~I Petrache, and R~M Suter.
\newblock X-ray structure determination of fully hydrated l alpha phase
  dipalmitoylphosphatidylcholine bilayers.
\newblock {\em Biophysical Journal}, 70(3):1419--1431, 1996.

\bibitem{nagle_tristram_2000}
J~F Nagle and S~Tristram-Nagle.
\newblock Structure of lipid bilayers.
\newblock {\em Biochimica and Biophysica Acta}, 1469(3):159–195, 2000.

\bibitem{kucerka_nagle_2006}
N~Kucerka, S~Tristram-Nagle, and J~F Nagle.
\newblock Closer look at structure of fully hydrated fluid phase dppc bilayers.
\newblock {\em Biophysical Journal}, 90(11):L83--L85, 2006.

\bibitem{kucerka_katsaras_2008}
N~Kucerka, J~F Nagle, J~N Sachs, S~E Feller, J~Pencer, A~Jackson, and
  J~Katsaras.
\newblock Lipid bilayer structure determined by the simultaneous analysis of
  neutron and x-ray scattering data.
\newblock {\em Biophysical Journal}, 95(5):2356–2367, 2008.

\end{thebibliography}
\end{document}